\documentclass{article}
\usepackage{emulateapj}
\usepackage{apjfonts}

\begin{document}

\newcommand\Msun {M_{\odot}\ }
\newcommand\Lsun {L_{\odot}\ }

\title{Deep Hubble Space Telescope Imaging of IC 1613 II. The Star Formation
History\footnotemark[1]}

\author{Evan D. Skillman}
\affil{Astronomy Department, University of Minnesota, Minneapolis,
MN 55455, USA; skillman@astro.umn.edu}

\author{Eline Tolstoy, Andrew A. Cole}
\affil{Kapteyn Institute, University of Groningen
PO Box 800, 9700AV Groningen, the Netherlands;
etolstoy@astro.rug.nl, cole@astro.rug.nl}

\author{Andrew E. Dolphin, Abhijit Saha}
\affil{National Optical Astronomy Observatories, PO Box 26372, Tucson,
AZ 85726, USA;
dolphin@noao.edu, saha@noao.edu}

\author{J. S. Gallagher}
\affil{Department of Astronomy, University of Wisconsin-Madison,
475 North Charter Street, Madison, WI 53706, USA;
jsg@astro.wisc.edu}

\author{R. C. Dohm-Palmer}
\affil{Astronomy Department, University of Minnesota, Minneapolis,
MN 55455, USA;
rdpalmer@astro.umn.edu}

\author{Mario Mateo}
\affil{University of Michigan, Department of Astronomy,
821 Dennison Building, Ann Arbor, MI 48109-1090, USA;
mateo@astro.lsa.umich.edu}

\footnotetext[1]{Based on observations with the NASA/ESA Hubble Space 
Telescope, obtained at the Space Telescope Science Institute, which is 
operated by the association of Universities for Research in Astronomy, 
Inc., under NASA contract NAS 5-26555. These observations are associated 
with proposal ID 7496.}


\begin{abstract}

We have taken deep images of an outlying field in the Local Group
dwarf irregular galaxy IC~1613 with the WFPC2 aboard the Hubble Space
Telescope in the standard broad-band F555W (V, 8~orbits) and  
F814W (I, 16~orbits) filters.  The photometry reaches to V $=$ 27.7
(M$_V$ $=$ $+$3.4) and I $=$ 27.1 (M$_I$ $=$ $+$2.8) at the 
50\% completeness level, the deepest to date for an isolated
dwarf irregular galaxy. 
We analyze the resulting color-magnitude diagram (CMD) and compare
it with CMDs created from theoretical stellar models using three
different methods to derive a star formation history (SFH) as well as
constrain the chemical evolution for IC~1613.
All three methods find an
enhanced star formation rate (SFR), at roughly the same magnitude
(factor of 3), over roughly the same period (from 3 to 6 Gyr ago).
Additionally, all three methods were driven to similar
age-metallicity relationships (AMR) which 
show an increase from [Fe/H] $\approx$ $-$1.3 at earliest
times to [Fe/H] $\approx$ $-$0.7 at present.
Good agreement is found between 
the AMR which is derived from the CMD 
analysis and that which can be inferred from the derived 
SFH at all but the earliest ages.
The agreement between the three models and the self-consistency of
the derived chemical enrichment history support 
the reality of the derived SFH of IC~1613 and, more generally, 
are supportive of the practice of 
constructing galaxy SFHs from CMDs.

A comparison of the newly observed outer field with an earlier
studied central field of IC~1613 shows that the SFR 
in the outer field has been significantly depressed during the last Gyr.
This implies that the optical scale length of the galaxy has
been decreasing with time and that comparison of galaxies at 
intermediate redshift with present day galaxies should take
this effect into account.
Comparing the CMD of the outer field of IC~1613 with CMDs of
Milky Way dSph companions, we find strong similarities between
IC~1613 and the more distant dSph companions (Carina, Fornax,
Leo I, and Leo II) in that all are dominated by star formation
at intermediate ages.  In particular, the SFH and AMR for IC~1613
and Leo I are indistinguishable.  This implies that dIrr galaxies 
cannot be distinguished from dSphs by their intermediate
age stellar populations.
This type of a SFH may also be evidence
for slower or suppressed early star formation in dwarf galaxies
due to photoionization after the reionization of the universe
by background radiation.
Assuming that IC~1613 is typical of a dIrr evolving in isolation, 
since most of the star formation occurs at intermediate
ages, these dwarf systems cannot be responsible for 
the fast chemical enrichment of the IGM which is seen at high redshift.
There is no evidence for any large amplitude bursts of star formation 
in IC~1613, and we find it highly unlikely that analogs of IC~1613 
have contributed to the excess of faint blue galaxies in existing 
galaxy redshift surveys.

\end{abstract}



\section{Introduction}

\subsection{Motivation}

The study of resolved galactic stellar populations provides a powerful tool
for determining the physical parameters of a galaxy such as the star 
formation history, the chemical composition and enrichment history, 
and dynamical history of the system.  By studying large numbers of
individual stars, and interpreting the observable 
parameters such as the morphology of the color-magnitude diagram (CMD),
kinematics, and metallicity it is possible to gain vital clues
concerning the evolution of galaxies (e.g., Hodge 1989; 
Olszewski, Suntzeff, \& Mateo 1996; Freeman \& Bland-Hawthorn 2002).
However, it is sobering to realize that much of our knowledge
of the detailed star formation histories (SFH) of galaxies beyond 1~Gyr ago
comes from the Milky Way and its nearest satellites (within $\approx$
200 kpc).  In the past, the limiting factors have been crowding,
resolution limits, and sensitivity for accurate stellar photometry from the ground.
Within the Local Group we always detect a `halo' of red stars (i.e., the 
`Baade sheet;' Baade 1963), which corresponds to the tip of the first 
ascent of the red giant branch (indicative of old stars).
For those galaxies with adequate observations, the presence of stars as old as 
those in globular clusters 
($>~10$ Gyr) have been detected via the presence of RR~Lyrae stars 
(e.g., IC~1613, Saha et al.\ 1992).  Nonetheless, quantitative 
information on the SFH at all epochs, and particularly for ages greater than 
1~Gyr, comes primarily from
the study of main sequence turnoff luminosities.
Thus, HST provides a unique opportunity to extend the sample beyond the
immediate vicinity and influence of large galaxies like the Milky Way
or M31.

In the nearby dwarf spheroidal satellites of our Galaxy, the
complex SFHs which had been indicated early on
(e.g., Mould \& Aaronson 1983) have recently been revealed
completely (see reviews by Gallagher \& Wyse 1994; Da Costa 1998;
Mateo 1998; Grebel 1999; van den Bergh 2000a).
While, on the one hand, all stars in the Draco, Sculptor and UMi dwarfs
are `old' ($t >$ 10 Gyr) as evidenced by a unique turnoff at $M_{V}
\approx +4.0$, the Carina, Fornax, Leo~II and Leo~I 
dwarfs all exhibit turnoff stars at luminosities ranging from 
$+$4.0 to $+$1.5, showing conclusively that star formation has 
occurred over several Gyr in these objects (e.g., Mighell 1990; 
Lee et al.\ 1993; Smecker-Hane et al.\ 1994; 
Mighell \& Rich 1996; Mighell 1997; Hurley-Keller, Mateo,
 \& Nemec 1998; Gallart et al.\ 1999a,b; Dolphin 2002). 
These objects {\it do} have some stars as old as in the MW globular
clusters, but the majority of stars are
of intermediate age.  These stunning results
fundamentally shape our understanding of galaxy formation and
evolution by clearly demonstrating the varieties of SFHs 
in nearby low mass dwarf spheroidals.
In the LMC and SMC, in addition to ongoing star formation, the
large range of field star ages attest to continued star 
formation at all epochs in the past (see Holtzman et al.\ 1999;
Hatzidimitriou 1999).  All this is in accord with the working hypothesis
that all galaxies began forming stars more than 10 Gyr ago, and that
star formation continues until the gas is used up or expelled.  
It is important to extend studies of this type to larger distances
because the Galactic satellite
galaxies constitute an extremely incomplete sample that is
environmentally biased by the proximity to the Galaxy.  The next
logical step is to examine the fossil record of star formation in
galaxies of various types and sizes, and to identify both
commonalities and differences in their SFHs.
One simple and still unanswered question is how dIrr and dSph
galaxies are related (Faber \& Lin 1983; Lin \& Faber 1983; 
Kormendy 1985; Dekel \& Silk 1986; Ferguson \& Binggeli 1994;
Gallagher \& Wyse 1994; Skillman \& Bender 1995; Mateo 1998;
Ferrara \& Tolstoy 2000: Skillman, C\^ot\'e, \& Miller 2003;
Grebel, Gallagher, \& Harbeck 2003).

The approach of gaining a detailed understanding of nearby galaxies
is a stepping stone to understanding
galaxy evolution, and thus provides a physical basis for understanding
observations of high red-shift galaxies and their implications for
cosmological models (Tolstoy 1999; Hopkins, Irwin, \& Connolly 2001).  
Often, dwarf galaxies are called upon to
solve a number of observationally motivated problems.
In the last decade, it has been suggested that dwarf galaxies
are responsible for enriching the intracluster medium
(Trentham 1994; but see Nath \& Chiba 1995) 
and the intergalactic medium (Nath \& Trentham 1997;
Murakami \& Babul 1999).  
There have been suggestions that the ``faint blue galaxy'' problem
(Tyson 1988; Koo \& Kron 1992; Colless et al.\  1993; 
Lilly 1993; Glazebrook et al.\ 1995) could be solved by a
bursting population of dwarf galaxies (Babul \& Rees 1992;
Babul \& Ferguson 1996).  
The luminosity functions of dwarf galaxies provide a severe constraint 
on the cold dark matter cosmologies (Klypin et al.\ 1999; 
Moore et al.\ 1999; Gnedin 2000a).  For all of these reasons, a 
better understanding of the SFHs of 
dwarf galaxies will help us to better understand the evolution
of all galaxies.

\subsection{The Dwarf Irregular Galaxy IC 1613}

IC~1613 is a member of our Local Group and one of the nearest 
gas rich irregular galaxies (for a review of the properties of
IC~1613 see van den Bergh 2000a).  Because of its proximity, IC~1613 
offers the opportunity to reconstruct a detailed 
SFH of a relatively isolated and non-interacting dwarf irregular galaxy
(as opposed to the significantly closer - and larger - Magellanic Clouds). 
IC~1613 also has very low
foreground and internal reddening, and very reliable distances from
Cepheid and RR~Lyr variable stars (Sandage 1971; Freedman 1988; Saha
et al. 1992; Dolphin et al. 2001b, hereafter D01).
It is located at a distance of 730~kpc (D01), approximately 250~kpc above the 
M31-Milky Way axis (Grebel 2001), and is included in the 
diffuse ``Local Group Cloud'' of galaxies by Mateo (1998). 

The physical parameters of IC~1613 were summarized in Cole et al.\ (1999,
 C99), which presented a WFPC2 stellar population study of a central field 
of IC~1613.  These properties are normal for an Im~V galaxy with  a
moderate luminosity ($M_V = -$15.2) and a small value of the maximum
amplitude of the rotation curve (V$_{max}$ $=$ 25 km s$^{-1}$,
Lake \& Skillman 1989).  Its star formation rate (SFR) of 
0.003~$\Msun$~yr$^{-1}$ given by Mateo (1998; cf., Hunter \& 
Gallagher 1986) is also normal for its type and luminosity. 
The current ISM oxygen abundance as 
determined from HII regions ranges from 12 $+$ log (O/H) $=$ 7.83
(Peimbert, Bohigas, \& Torres-Peimbert 1988), which corresponds
to 14\% of the solar oxygen abundance as determined by
Allende Prieto, Lambert \& Asplund (2002), to 12 $+$ log (O/H) $=$ 7.62
(Lee, Grebel, \& Hodge 2003) corresponding to 8.5\% 
of the solar value.  This is slightly less than that
of the SMC, and normal for a galaxy of its luminosity 
(Skillman et al.\ 1989).  
This combination of proximity and normalcy makes IC~1613
one of the best laboratories to study the properties of a dwarf star-forming
galaxy that is relatively isolated (as is typical for Im~V galaxies). 

The overall structure of IC~1613 also is typical of an Im~V galaxy. 
While the central field observed by C99 is relatively quiescent, at 
slightly larger radii OB associations, HII regions and HI shells are found 
(e.g., Hodge 1978; Lake \& Skillman 1989; Price, Mason, \& Gullixson 1990; 
Georgiev et al.\ 1999; Valdez-Guti\'errez et al.\ 2001).
Over the last few years there have been several papers on the resolved
stellar population of IC~1613 based on HST and high quality 
ground-based observations. C99 found IC~1613 to be a 
smoothly evolving galaxy with a relatively constant SFR over its 
lifetime.  A search for bona fide star clusters (Wyder, Hodge \& Cole 2000), 
of which IC~1613 has a marked lack in comparison with the SMC
(Baade 1963; van den Bergh 1979), turned up only one 10~Myr old cluster 
in images taken with the WIYN telescope.  The relative lack of
star clusters could be taken as evidence that IC~1613 has been 
evolving in isolation, free from galaxy-galaxy interactions which
may be responsible for triggering most cluster formation 
(cf.\ van den Bergh 2000b).   

Borissova et al.\ (2000) studied the distribution of luminous cool
stars from J and  K-band imaging 
obtained with the 2.1-m telescope at the San Pedro Martir
Observatory. They found young supergiants in the central northeast
star forming region and AGB stars covering a wide range in age in all 
of their inner galaxy fields.  Albert, Demers, 
\& Kunkel (2000) surveyed IC~1613 for C and M stars with a wide field 
camera on the Dupont Telescope. This extension of the original 
Cook, Aaronson, \& Norris (1986) C star survey shows that 
the old stellar population spreads well beyond the regions where 
star formation  currently is active. The C/M star ratio is typical 
for a galaxy of this luminosity (cf., Battinelli \& Demers 2000; 
Demers \& Battinelli 2002), which 
can be taken as prima facie evidence against large variations in the
SFR at intermediate ages.

Variable star investigations have offered an additional dimension 
to studies of the stellar populations in IC~1613. A variety of types 
of variable stars are seen (e.g., Mantegazza et al.\ 2001), 
including the recent discovery of a luminous Mira (Kurtev et al.\ 2001).  
This galaxy has a reasonable number of Cepheids and has been used 
to study the period-luminosity relationship (e.g., 
Sandage, Bell, \& Tripico 1999; Antonello, Fugazza, \& Mantegazza 2000).
The presence of a spectrum of pulsating variables extending from long to 
short periods (see D01) is consistent with star formation having been active 
in IC~1613 since $>$10~Gyr in the past.

Here we present a deep color-magnitude diagram (CMD)
for an outer field in IC~1613
from long integrations with WFPC2 in the F555W and F814W
filters. We derive the first detailed model of the SFH for this field from
CMD model fitting analysis, reaching back to the oldest epochs.
Because these data consist of such deep integrations, we detect stars
further down the main sequence than has previously been possible for
an irregular galaxy beyond the Magellanic Clouds. These data thus
have the quality to allow comparisons between
different approaches to CMD analysis.  We present three independent
derivations of the SFH of our IC~1613 outer field using the
techniques of Tolstoy \& Saha (1996); Dolphin (1997, 2002) and Cole
(2003; cf.\ Smecker-Hane et al.\ 2002).

\section{The Data}

\begin{figure}
\plotone{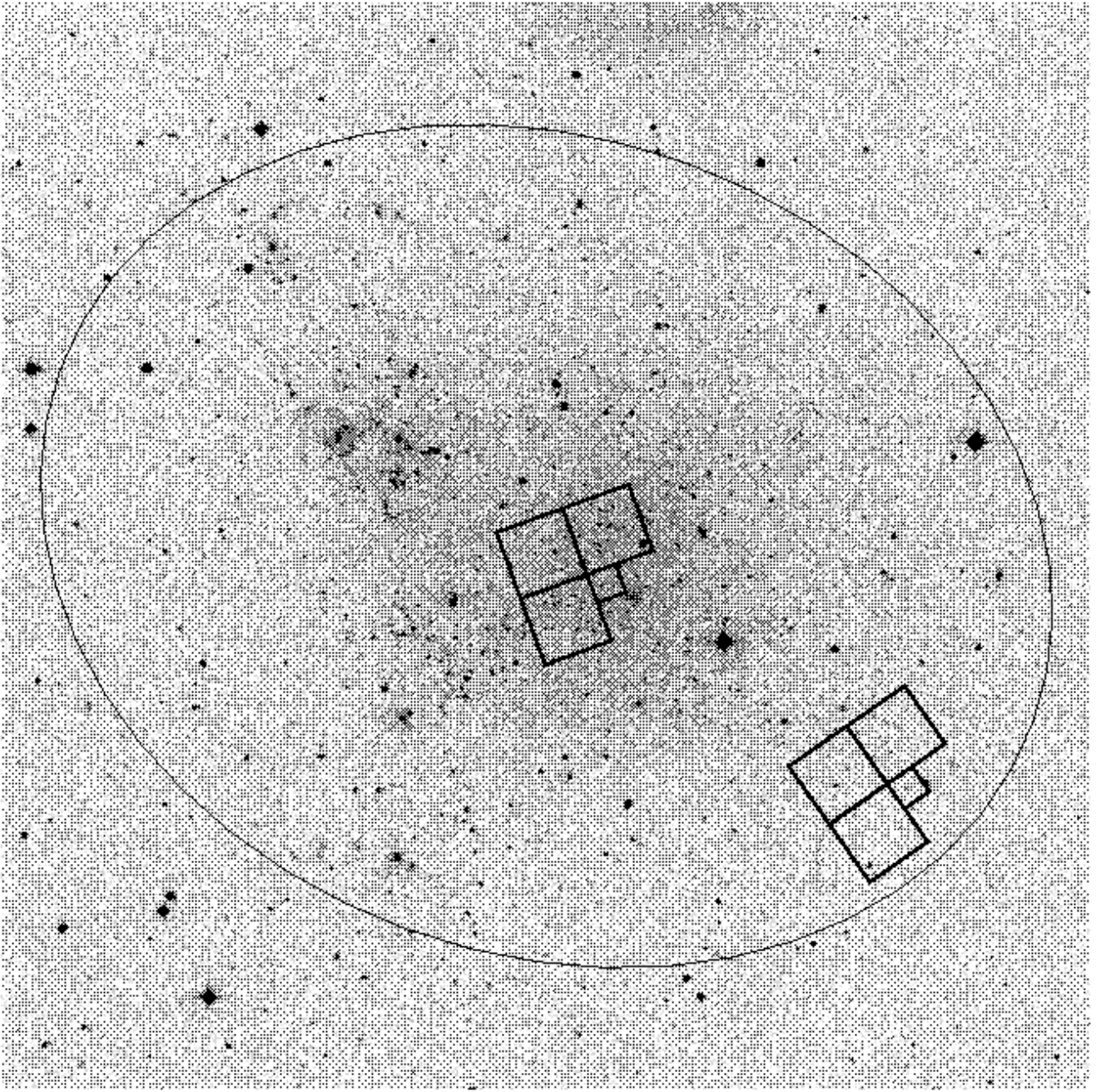}
\caption{ Positions of the (earlier) ``central'' and (new) ``outer'' HST WFPC2 
observations overlaid on the DSS image of the IC~1613.
The image is 20$\arcmin$ on a side.  The ellipse shows the
extent of the carbon stars identified by Albert et al.\ (2000),
and is included to show where the red stars in IC~1613 begin to 
dominate the background.
}
\label{footprint}
\end{figure}

\begin{figure}
\plotone{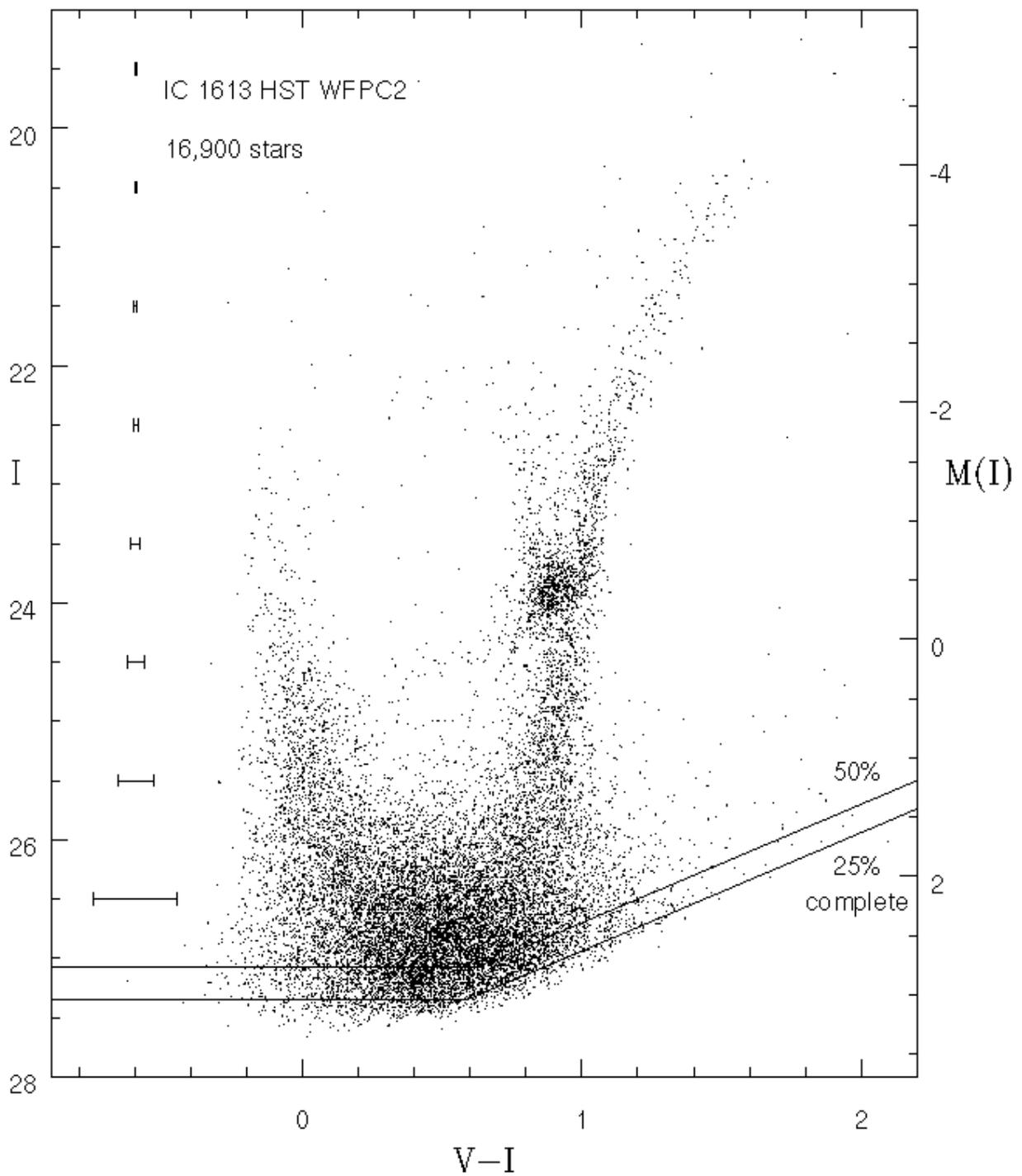}							     
\caption{ CMD of IC~1613 derived from HST WFPC2 observations.
The error bars on the left of the diagram show representative
photometric errors.  The two lines across the bottom show the 
incompleteness limits at the 25\% and 50\% levels as derived
from false star tests.  The right hand scale gives the absolute
magnitudes corresponding to a distance of 730 kpc as derived 
from D01.)
}
\label{cmd}
\end{figure}

These data have been described by D01, who used the HST single orbit 
integrations to identify and photometer variable stars 
such as RR~Lyrae and short period Cepheids.
As shown in Figure~\ref{footprint}, an overlay of the WFPC2 footprint
on the STScI Digitized Sky Survey\footnotemark , 
the field center is located 7.4$\arcmin$ SW from the dynamical 
center (Lake \& Skillman 1989, a projected distance of 1.6 kpc) 
and 6.7$\arcmin$ (1.4 kpc) 
SW from the ``central'' field studied by C99.  The position of
this field was chosen to minimize crowding and contamination by
bright young stars and yet still produce a reasonably large 
number of stars.  
Altogether, 8 orbits produced 16 1200s V band (F555W) images and 
16 orbits produced 32 1200s I band (F814W) images.  These images
were obtained at four slightly different ``dithered'' positions, 
and so four deep images were made in each band.  Photometry was 
conducted using the {\it multiphot} routine of HSTPHOT 
(Dolphin 2000a) applying the charge transfer loss corrections
as described in Dolphin (2000b). 
In Figure~\ref{cmd} we present the final CMD which is very similar
in appearance to Figure 2 of D01. 
Representative photometric error bars are plotted on the left in 
Figure~\ref{cmd} and the absolute magnitude scale on the right 
assumes a distance of 730 kpc as derived by D01.
\footnotetext{The Digitized Sky Surveys were produced at the Space Telescope
Science Institute under U.S.  Government grant NAG W-2166. The images of
these surveys are based on photographic data obtained using the Oschin
Schmidt Telescope on Palomar Mountain and the UK Schmidt Telescope.
The plates were processed into the present compressed digital form with
the permission of these institutions.}

The CMD displayed in  Figure~\ref{cmd}, shows a relatively narrow 
and well defined
red giant branch (RGB), a well populated red clump (RC), a faint
horizontal branch (HB) which extends out from below the RC, and a main 
sequence (MS) that reaches above the RC, but is not nearly as 
well populated as in more central field of IC 1613 of C99.  Our photometry
follows the MS roughly two magnitudes below the HB.  The goal of
this program was to reach to the oldest possible MS turn-off 
populations.  In fact, the data do not reach as deeply as was 
originally proposed, and the limit of our data is roughly 0.5 
magnitudes brighter than our proposed goal (V $\approx$ 27.6).  
This difference is likely due to overly optimistic results from
the WFPC2 exposure time calculator (Li Causi, De Marchi,  \& 
Paresce  2002) and not properly accounting
for the higher background due to IC~1613's proximity to the 
ecliptic (ecliptic latitude $=$ $-$4.4$\arcdeg$). 
 
\begin{figure}
\plotone{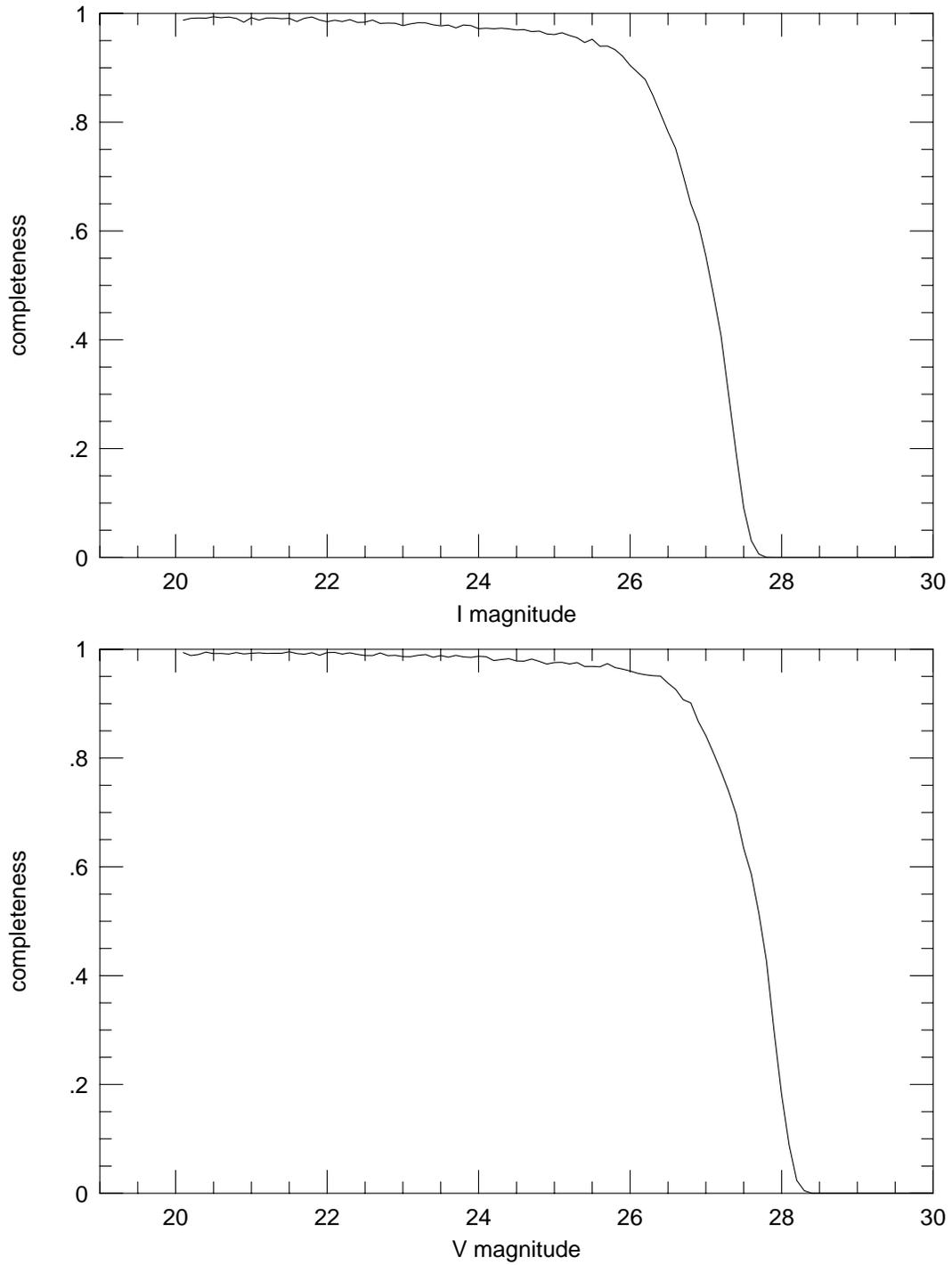}
\caption{ Incompleteness of IC~1613 V (lower panel) and I (upper
panel) photometry as determined from false star tests.
}
\label{incom}
\end{figure}

\begin{figure}
\plotone{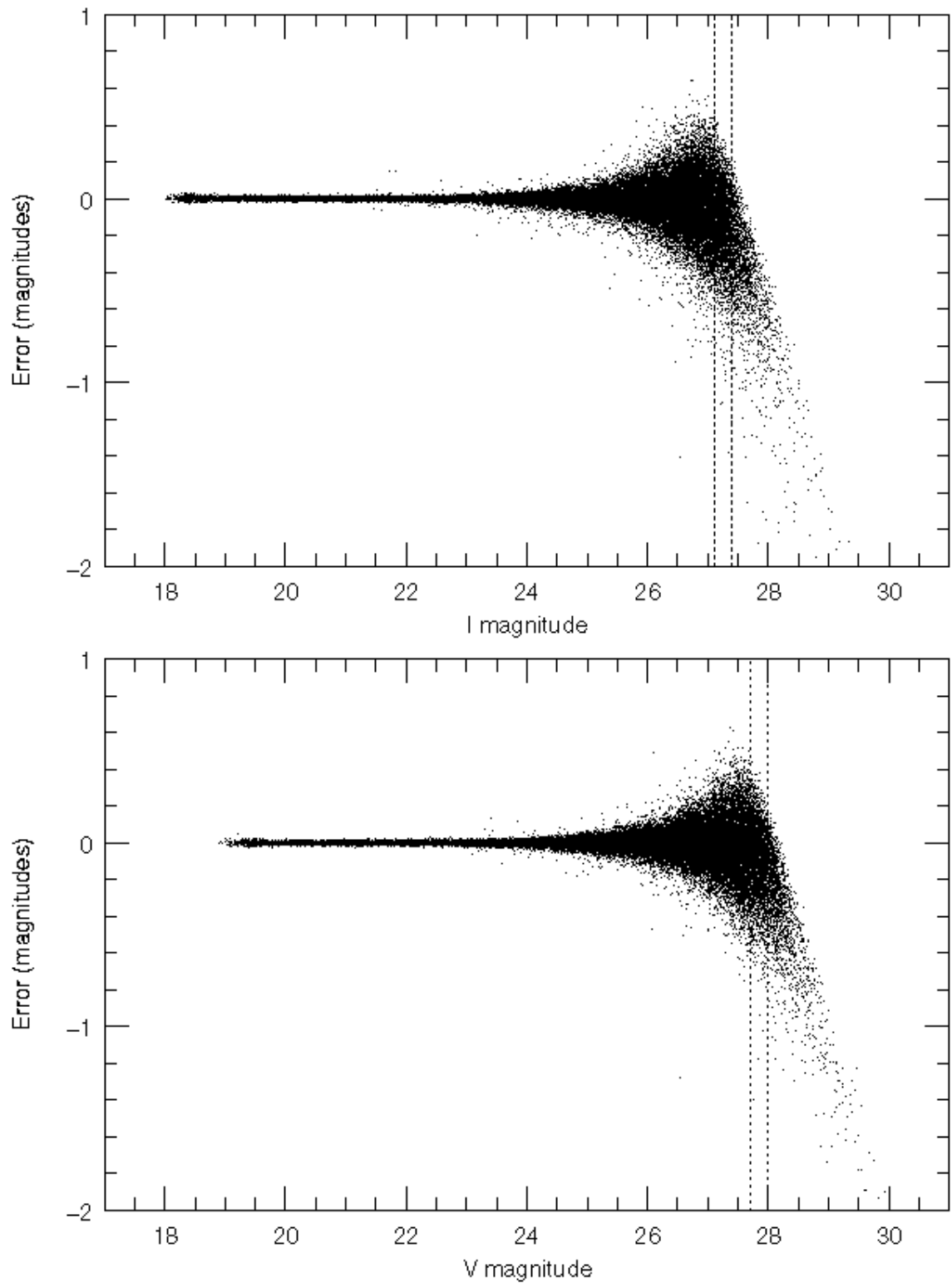}
\caption{ Errors in recovered magnitude as a function of input 
magnitude for the false star tests.  The vertical dotted lines
represent the 50\% and 20\% completeness limits of the observations.
}
\label{fake}
\end{figure}

The incompleteness, or the likelihood of detecting a star versus its
magnitude in our CMD, has been calculated using the precepts laid out 
in Dolphin (2002).  This is a critical step 
which is necessary for the accurate interpretation of a CMD,  and 
results are plotted in Figure~\ref{incom}. 
The 25\% and 50\% incompleteness limits determined from Figure~\ref{incom}
have been plotted in Figure~\ref{cmd}.
Since this is an uncrowded field, the increase 
in both error and incompleteness at faint magnitudes is primarily due 
to low S/N, and not crowding (i.e., we have not yet hit the confusion 
limit).  In Figure~\ref{fake} we have plotted the error in the recovered
magnitude as a function of input magnitudes for the false star tests.
The main purpose of Figure~\ref{fake} is to remind the reader of the
rather substantial photometric errors than can occur well before 
incompleteness becomes significant.

\begin{figure}
\plotone{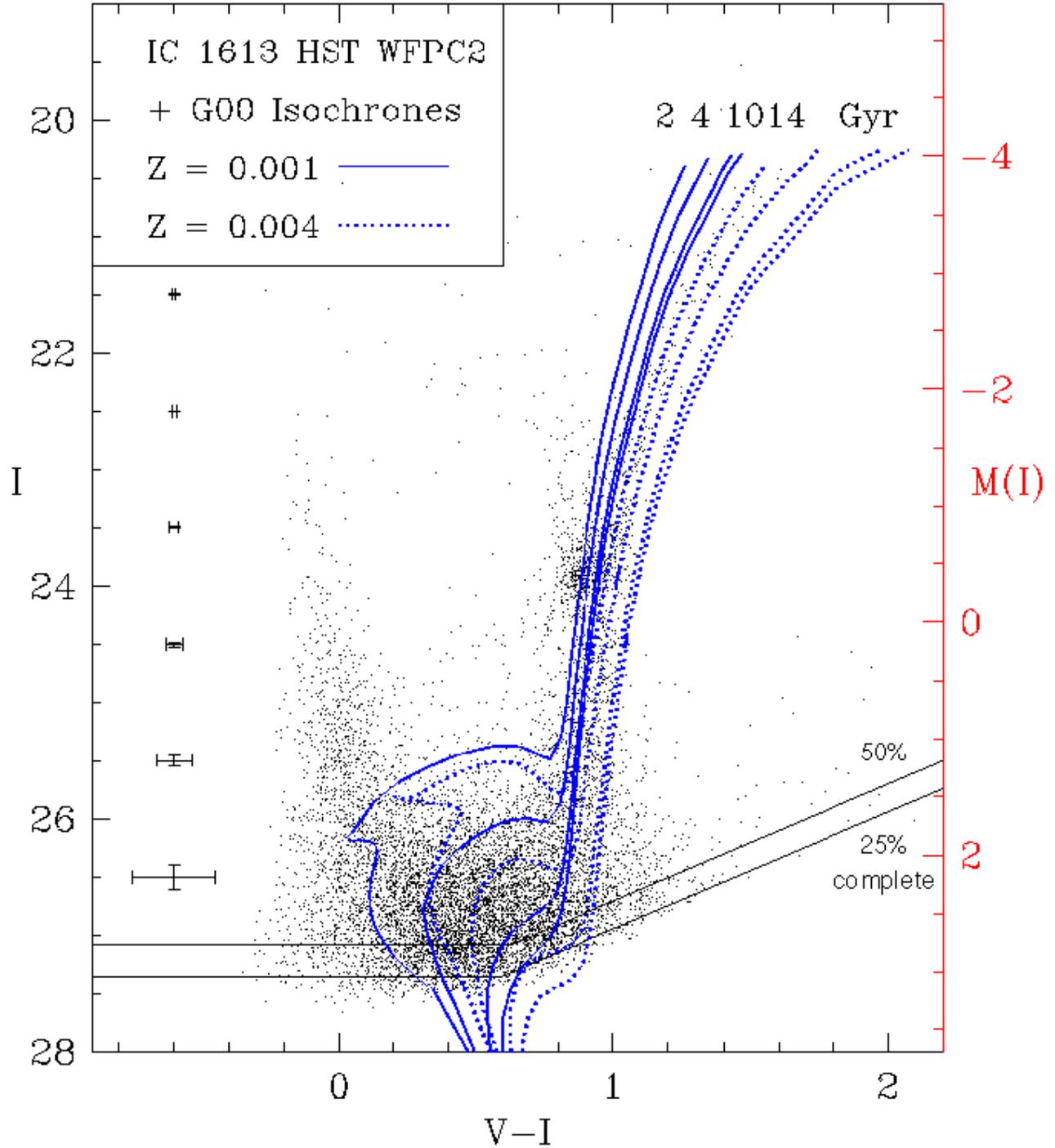}							     
\caption{ CMD of IC~1613 derived from HST WFPC2 observations
as shown in Figure~\ref{cmd}.
Isochrones for a metallicity of $Z$ $=$ 0.001 (solid lines) and
$Z$ $=$ 0.004 (dotted lines) and ages of 2, 4, 
10, and 14 Gyr from Girardi et al.\ (2000) have been added in
order to show the limitations of the observations in terms of
MS ages.  MS turnoffs
back to intermediate ages ($\sim$5 Gyr) are well represented in
the observations, but the oldest MS turnoffs ($\sim$10 Gyr) fall
below the 50\% completeness limit and 
are not represented.  Thus, constraints on the oldest populations
will need to come from the evolved stars.
}
\label{cmd2}
\end{figure}

In Figure~\ref{cmd2} we have overplotted theoretical isochrones on
top of the CMD shown in Figure~\ref{cmd}.  The isochrones were 
calculated for a metallicities of $Z$ $=$ 0.001 and 0.004 and ages of 2, 4,
10, and 14 Gyr by Girardi et al.\ (2000). The choice of the low
metallicity (5\% of solar) was driven mostly by the position of
the RGB, but the position of the oldest isochrone to the left
of the reddest RGB stars indicates that even the oldest stars
are slightly more metal rich than this.  However, the positions of the 
higher metallicity isochrones show that the lower metallicity 
is a relatively good fit to the RGB stars. 
The main reason for overlaying the isochrones here
is to demonstrate the coverage and limitations of the observations
with respect to the MS turnoffs.  MS turnoffs
back to intermediate ages ($\sim$5 Gyr) are well represented in
the observations.  However, the oldest MS turnoffs ($\sim$10 Gyr) 
fall below the 50\% completeness limit, and thus, are not well 
represented in the observations.  As a result, in constructing SFHs,
information on the oldest populations will need to come from 
evolved stars.

\section{The Models}

In this section and the next we describe an exercise of comparing 
different modeling techniques in order to determine the SFH of IC~1613
from the present observations.  The interpretation of the derived
SFH begins in \S 5.

One concern in the field of SFH construction is the uniqueness of the 
solution (e.g., Tolstoy \& Saha 1996; Aparicio et al.\ 1996; 
Dolphin 1997; 2002; 
Gallart et al.\ 1999a; Holtzman et al.\ 1999; Hernandez et al.\ 1999; 
Harris \& Zaritsky 2001; Skillman \& Gallart 2003). 
As higher quality data are obtained and stellar evolution models continue
to improve, observational constraints on the
derived SFHs become increasingly stringent,
and uniqueness becomes less of a problem.  However, as one attempts
this type of work at ever increasing distances, one is faced with
the challenge of always trying to do more with less.  Given the
relatively high quality of the present dataset, we judged that it would
be valuable to conduct three independent SFH derivations.
By comparing the results of three independent
approaches, we can ascertain which features of the SFHs
are technique independent and which are not.  We can
also compare the differences between the three results to the 
sizes of the errorbars and judge the quality
of the error assessment in each technique.

Here follow three independent approaches to the analysis of the same 
observational data.  We present the results from
each exercise followed by a comparison of all three.   
In each section we only briefly describe each 
technique; references to other papers describing the details of the 
techniques are given.
Our aim here is to investigate whether these relatively high
quality data drive all three investigations to similar results
(are the results robust?), or if different choices in technique
result in significant differences in interpretation (are the
derived SFHs plagued by systematic 
uncertainties?).  We are not attempting to determine which of
the three methods is best.  That debate has both philosophical and
technical concerns and is better done with comparisons
over a large range in synthetic and real observations.  Nor
are we claiming that if two different methods give similar results
that differences in methodology do not matter. 
This issue would involve discussions of other
methods for deriving SFHs and lies beyond the scope of this paper.

One way to characterize the three different methods is the degree
to which the solution is constrained (e.g., by assumptions of certain
values or by limitations on the number of variables).  In the Tolstoy 
method, statistical measurements are used to guide choices between 
prescribed solutions.
In the Cole method, a chemical enrichment law is assumed, and the
SFR is solved for as a function of time.  In the
Dolphin method, both the metallicity and SFR are
solved for as a function of time.  Note, however, that there are 
several other differences including: interpolation of the stellar 
evolution models in metallicity, allowing for the presence of binary stars, 
assuming or solving for the distance and/or reddening, etc.

It is important to emphasize that we are {\it not} testing the 
quality of the stellar evolution models or the suitability of the
stellar atmospheres used to calculate the individual star's luminosities
and colors.  All three techniques use the same evolution models as
their basis, so this cannot be tested here.  Single age and 
metallicity samples (star clusters) provide more suitable data
for testing and tuning of the stellar evolution models (e.g., 
Chiosi, Bertelli, \& Bressan 1992; 
Carraro et al.\ 1994; Bragaglia et al.\ 2001; 
Barmina, Girardi, \& Chiosi 2002; 
and references therein).
 
\subsection{The Tolstoy approach}

Following updated precepts first laid out in Tolstoy
(1996), Tolstoy \& Saha (1996), and Tolstoy et al.\ (1998), a model has
been built up of the SFH of IC~1613.  
The results
from stellar evolution codes are used directly as input and there is
no interpolation between different metallicity tracks.  Also, the
photometric data are not binned and the effects of binaries are not
directly accounted for.  The present
technique is not automated and relies on human judgement to limit the
search to various solutions.  
Using an  archive of stellar evolution tracks converted to
observables and Monte-Carlo simulation techniques, 
a model CMD is built that best matches the observed CMD, and from this 
a plausible SFH for IC~1613 is inferred.  
The Padua stellar evolution tracks (Girardi et al.\ 2000) were 
used, and then, following Tolstoy (1996), the
temperature and luminosities provided by the Padua group were then
converted to observed I, V$-$I magnitudes using Kurucz models (Kurucz
1993). This resulted in slightly different observed magnitudes and
colors to the Padua isochrones, 
but very similar to the new Yonsei-Yale isochrones 
(Yi et al.\ 2001; 2003).

A simple premise is used as a starting point - 
that of constant star formation at a constant metallicity throughout 
the history of this galaxy.  Successive iterations then
investigate where this model fails and what needs to be changed to
correct these failings.  Note that because the stellar evolution 
models are not interpolated in metallicity, the degree of
failure must be large enough to justify the ``jump'' to the next
set of models.

For the young stellar population, the best fitting tracks were those of
metallicity Z=0.004 (or [Fe/H] $= -0.7$), or 20\% solar.  This is
fairly robust, as the upper part of the CMD (I$<$23.5) contains an
obvious Z=0.004 blue loop population (rising vertically out of the red
clump, I$<$23.5 and V$-$I$\sim$0.8), and not the distinctive lower
metallicity blue loop population which would populate a 
diagonal strip across the CMD from the red clump toward the upper main
sequence (e.g., Sextans A; Dohm-Palmer et al. 1997, 2002).  These
stars have ages $\leq$1~Gyr.  This metallicity is in agreement with 
the nebular H~II region oxygen abundances of 12 $+$ log(O/H) $=$ 7.86
reported by Skillman, Kennicutt, \& Hodge (1989).  If one assumes
constant [Fe/O] $=$ 0, then this is equivalent to [Fe/H]$= -0.8 \pm 0.2$,
however, taking the new results of Lee et al.\ (2003) gives a 
slightly more metal poor equivalent [Fe/H] of $-$1.07. 
This is more metal rich than previous estimates of the metallicity
of the old population based on the RGB ([Fe/H]$= -1.3 \pm 0.2$) 
from Lee et al.\ (1993), and thus, perhaps indicative of enrichment.

The best understood, most accurate SFR indicator is
the main sequence, and this forms the backbone of the constraints on
the model.  These data
allow us to follow the main sequence down to about 8$-$9~Gyr old
turnoff ages, but our sensitivity to all but large variations in the
SFRs is very poor $\ge$ 6~Gyr ago.  There is no
evidence for any {\it large} (i.e., factor of 5 or greater) variation 
in the SFR in this
field of IC~1613 over the last 6$-$8~Gyr. We find that the stellar
density in the lower part of the CMD (I=27, 0.2 $>$ V$-$I $<$ 0.8)
requires an enhancement (a factor 2$-$3) in the global SFR
between 3 and 7 Gyr ago. The lower limit of this age
range (3~Gyr) is determined by the relative lack of any enhancement in
stellar density higher up the main sequence, and the upper limit
(7~Gyr) is determined, arguably somewhat insecurely, by 
the fact that modeling
the SFH thus far has already over-populated the RC
and RGB and covered the entire color range of the RGB as well.  

A small population older than 7~Gyr is required by the
small protuberance on the blue side of the RC - this requires a lower
metallicity (Z=0.001), old ($>$~10~Gyr) RC.  The presence of an old
($\sim$ 10 Gyr) stellar population is also required by
the detection of RR~Lyr variable stars in IC~1613 in general
(Saha et al.\ 1992) and this field in particular (D01).

Some researchers use the HB as an additional constraint in
deriving SFHs, but it is not used here.
The presence and morphology of the HB in the CMD provides very limited 
insight into the SFH at this epoch due to the second 
parameter effect (Fusi Pecci et al.\ 1992; 1993; Bellazzini et al.\ 2001; 
Catelan et al.\ 2001; and references therein).  The HB population is 
certainly a small fraction of the stars in the CMD.  It might be that 
the slight enhancement in the SFR at $\sim$~1~Gyr is due to an over 
density in the MS due to a blue horizontal branch population. This was 
also hypothesized by C99 for the central field of IC~1613.

The majority of star formation in this field appears to have occurred
in the last 5$-$7~Gyr ago, with a peak between 4$-$6 Gyr ago. If 
a lower metallicity (Z=0.001) is invoked at 6~Gyr ago, this would make the
peak in the SFR diminish somewhat and bring the SFH
closer to a constant rate over time.  Additional information is
required to select either one of these scenarios.
This is the same dichotomy we found in modeling the Pegasus dI  
(Gallagher et al. 1998).

\begin{figure}		
\plotone{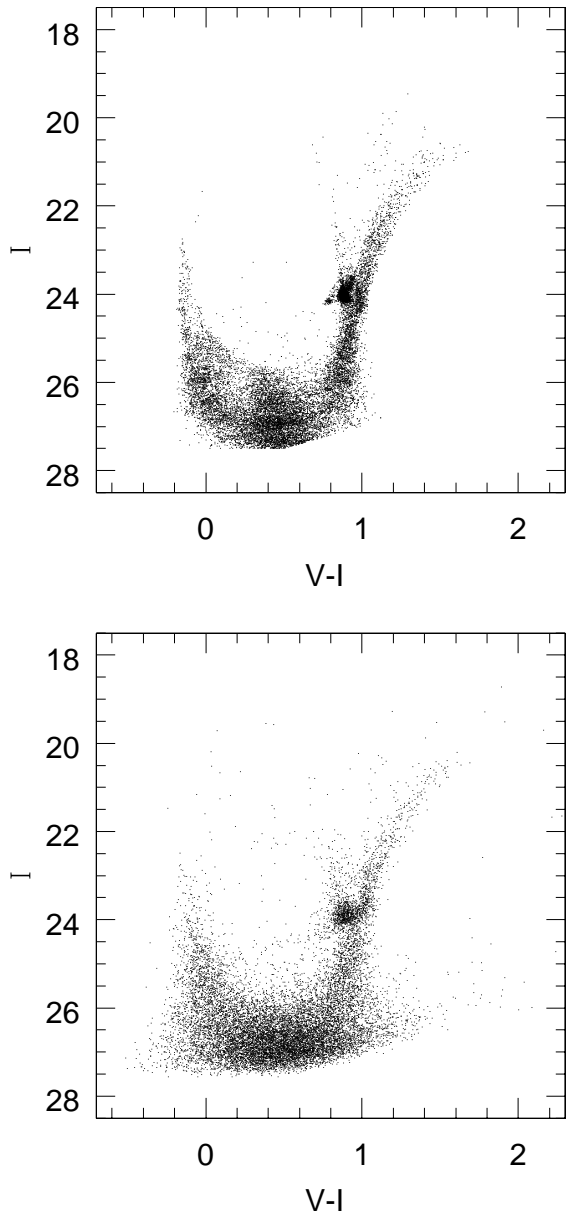}						     
\caption{Synthetic CMD of the best fit model to IC~1613 derived
via the Tolstoy method (upper panel) compared to the observations
displayed on an identical scale (lower panel).  The most notable
difference is the larger width of the RGB in the model.
}
\label{etmod}
\end{figure}

\begin{figure}	
\plotone{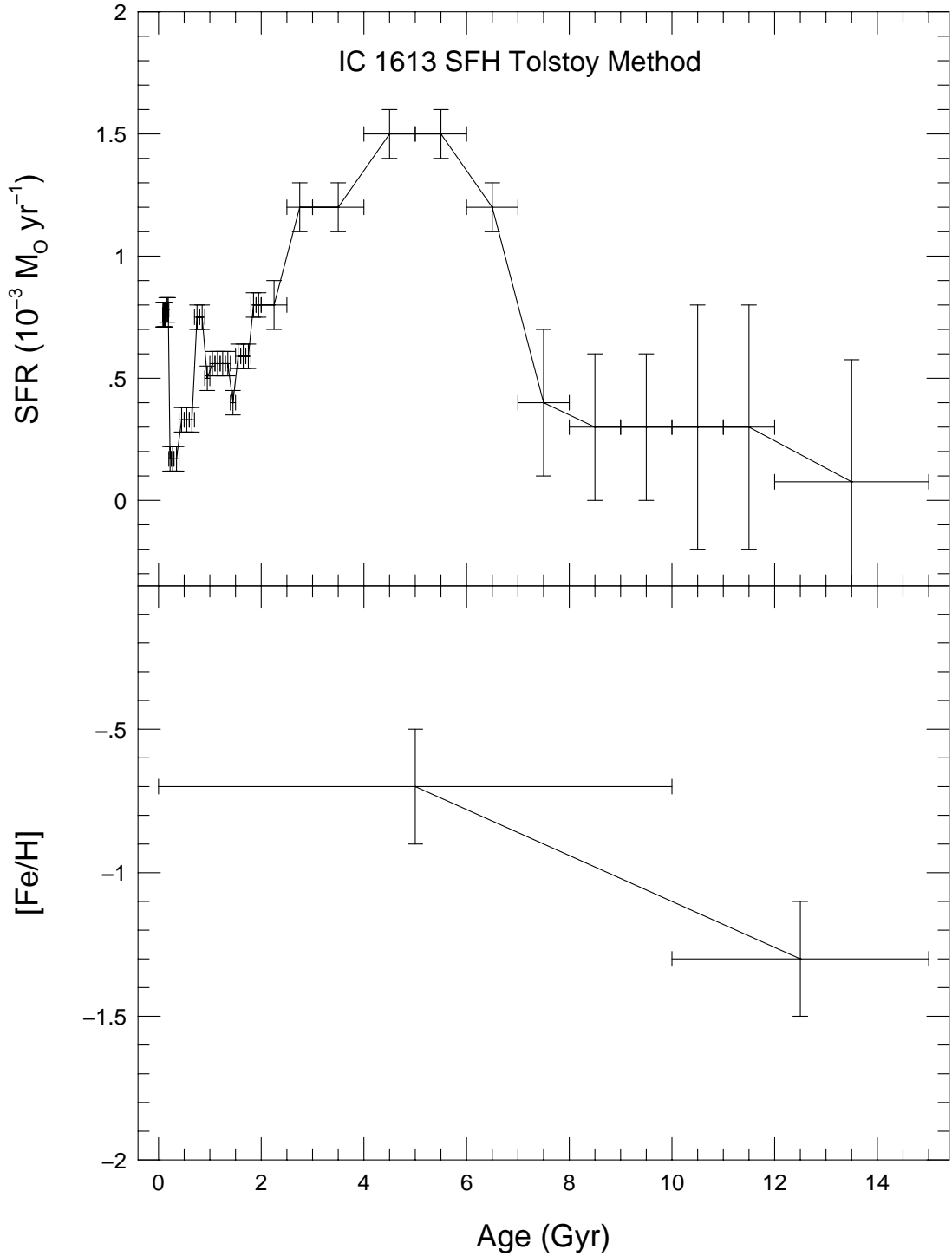}							     
\caption{Plot of most likely SFH and metallicity evolution  
of IC~1613 derived via Tolstoy method.  See text for discussion
of derivation of error bars.  Note the relatively enhanced
SFR between 2 and 6 Gyr ago.
}
\label{et}
\end{figure}

The model CMD which comes out of the above process and best matches
the observed CMD is shown in Figure~\ref{etmod}. This model is the
result of the SFH and the variation in metallicity 
with time shown in Figure~\ref{et}.
The errors in Figure~\ref{et} are, as discussed in Tolstoy et al.\ (1998),
representative error bars that come from the (large) sample of models
that have been created to find the best match to the SFH of IC~1613.
They give an indication of how far the SFR in any
bin can vary before the fit gets significantly worse.  This is 
complicated by the fact that a bad fit in one bin can sometimes
be compensated by changes in other bins.  Thus, the time bins are not
independent of each other, and this is quite difficult to account
for when determining the errors. 
Note that the errors in Figure~\ref{et} are only representative of 
limitations of the models used. They are not errors calculated from 
the models or any comparison with the data.

\subsection{The Cole approach}

This approach consists of determining the SFH 
by using a downhill simplex
algorithm with simulated annealing to minimize the 
difference between the observed and synthetic Hess diagrams.
The distance, reddening, and age-metallicity relation were
held fixed; the fit parameters were the SFRs
in 13 logarithmically-spaced age bins.  A $\chi ^2$ statistic
was used to measure the goodness-of-fit of the models.
The full details of the code will be specified in Cole (2003).  

The data were binned by 0.05 mag in (V$-$I) and 0.1 mag in I
to create the Hess diagram.  The models were created by using
a Monte Carlo technique to sample from theoretical stellar 
isochrones of Girardi et al.\ (2000).  The isochrones were
interpolated to arbitrary combinations of age and metallicity
in order to match the data.  Artificial star test results were
used to simulate the noise and completeness properties of the
data, so that a direct comparison could be made.  The stellar
IMF was taken from Kroupa, Tout \& Gilmore (1993), and binary
star populations were included following Duquennoy \& Mayor (1991)
and Mazeh et al.\ (1992).  Under
this prescription, 33\% of stars are single, 18\% are close 
binaries, and 49\% are wide binaries.  Close binaries have
secondary masses drawn from a flat mass function, while the 
secondaries of wide binaries are drawn from the Kroupa et al.\
(1993) IMF.  The adopted distance modulus (m$-$M)$_0$ = 24.31
and reddening E(B$-$V) = 0.02 were taken from the literature
(D01 and references therein).

The age-metallicity relation (AMR) was based on previous work
(see C99 and references therein), and iteratively modified
based on the colors of the red clump, RGB, and blue loop stars.
The AMR can be approximated by a constant value [Fe/H] = $-$1.4
for ages older than 10 Gyr, increasing linearly with time to
[Fe/H] = $-$0.7 at the present time.  No dispersion in abundance
at a given age is accounted for.

\begin{figure}	
\plotone{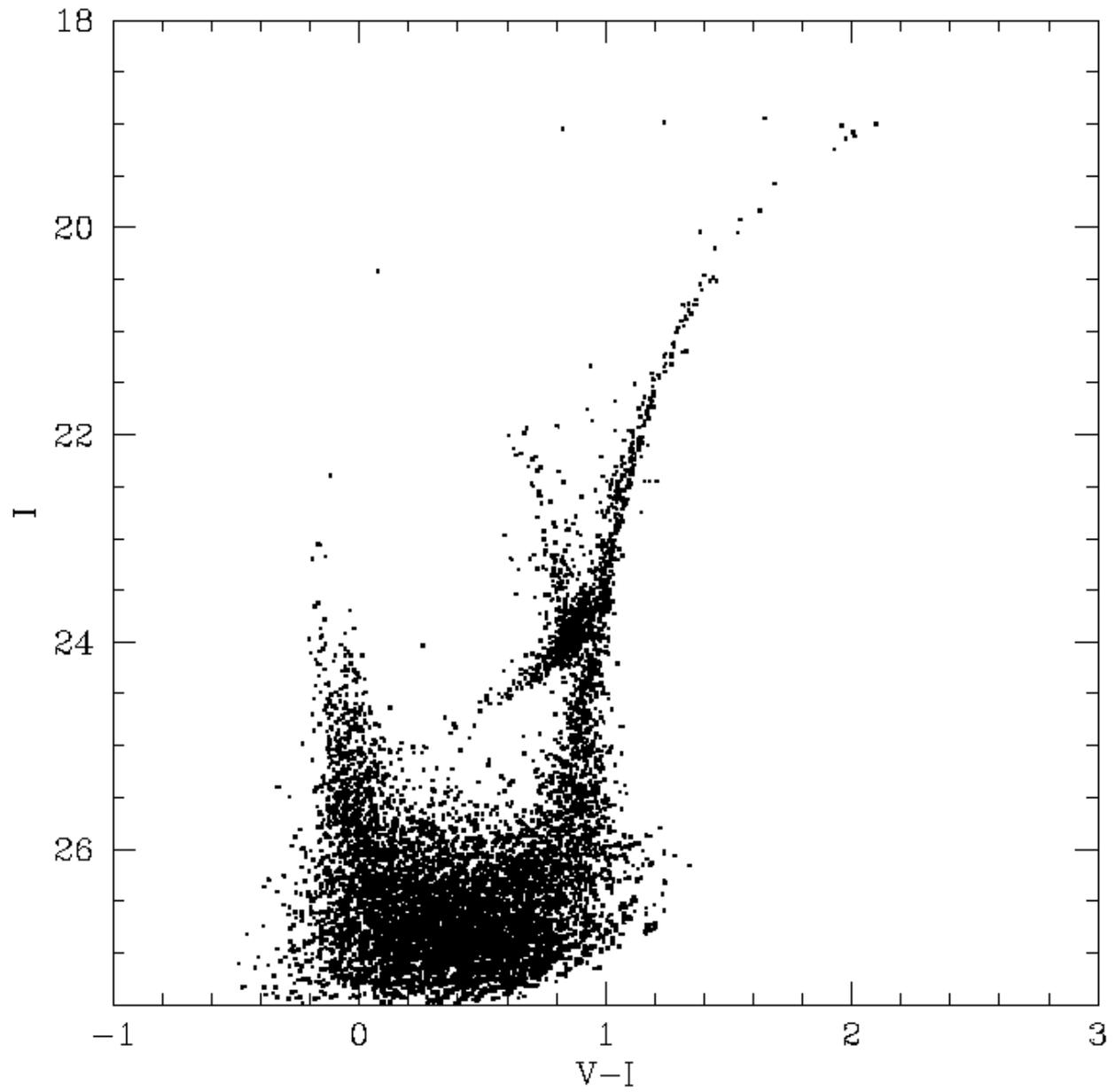}						     
\caption{ Synthetic CMD of the best fit model to IC~1613 derived
via the Cole method.  Note the very narrow RGB and the overpopulated
red HB when compared to the observational data shown in 
Figure~\ref{cmd}.
}
\label{aacsyncmd}
\end{figure}

\begin{figure}	
\plotone{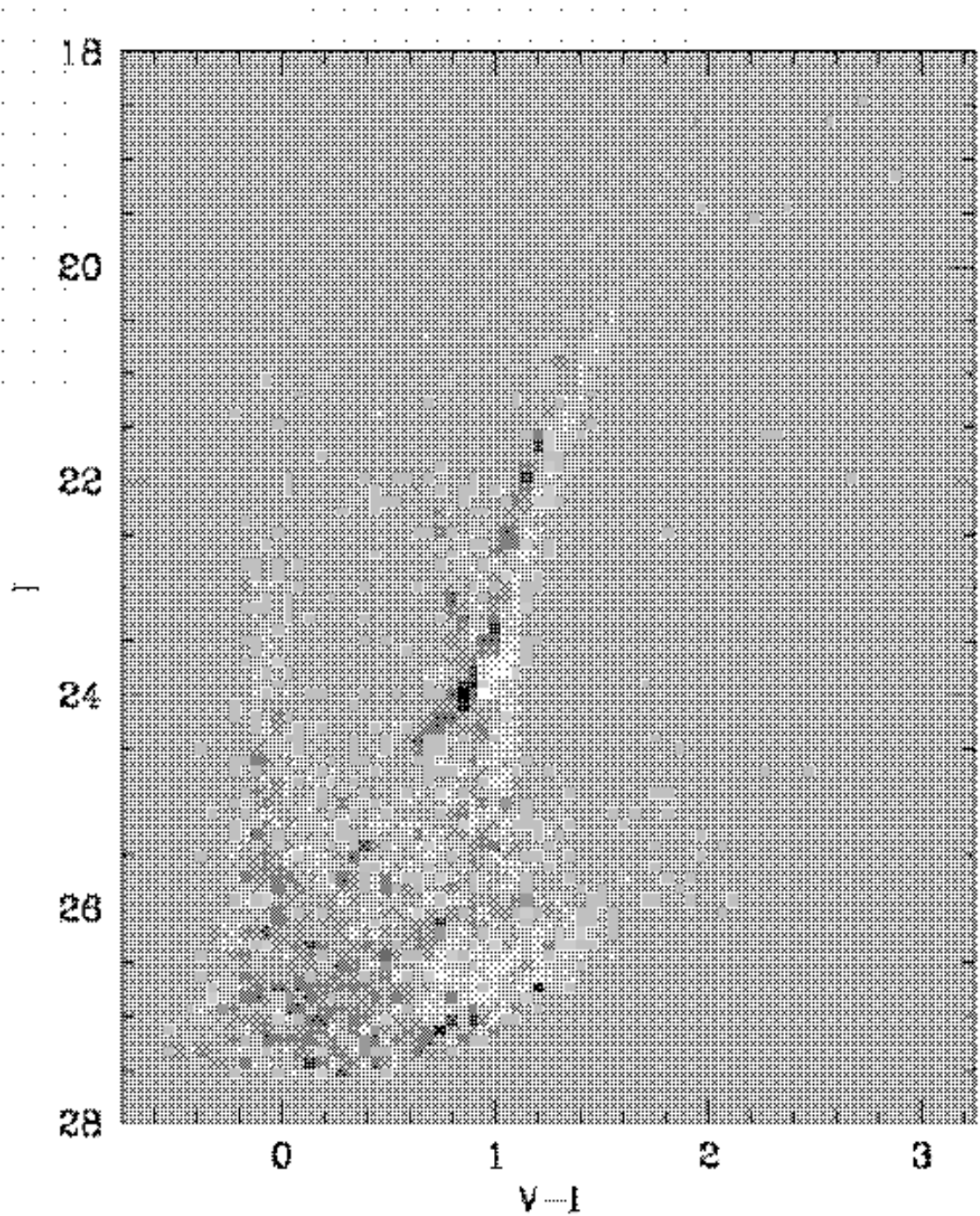}						     
\caption{A differential Hess diagram, comparing the best fit model to 
IC~1613 derived via the Cole method to the observational data when
binned identically.  White corresponds to a 5$\sigma$
excess in the data, and black to a 5$\sigma$ excess in the
model.  Note the excess of red HB stars and stars on the blue side 
of the RGB in the model. 
}
\label{aacdiffcmd}
\end{figure}

The synthetic CMD that results from this procedure is shown
in Figure~\ref{aacsyncmd}.
The general morphology of the CMD is similar to the 
observed CMD.  However, the upper two magnitudes of the
red giant branch are narrower than the observed RGB, and 
the synthetic CMD seems to lack upper main-sequence and bright
blue loop stars.  These effects are likely attributable to the
assumed AMR, which does not allow for any
abundance dispersion at a given age.  The model also overproduces
red horizontal branch stars, which could be a result of 
inadequacy in the isochrone set.  The overall quality of the
fit to the SFH is shown by the differential
Hess diagram in Figure~\ref{aacdiffcmd}.  
The figure shows the 
residuals from subtracting the synthetic Hess diagram from 
the data, scaled by the Poisson counting statistics in each
color-magnitude bin so that white corresponds to a 5$\sigma$
excess in the data, and black to a 5$\sigma$ excess in the
model.  In only a few pixels does the difference
exceed 4$\sigma$, most prominently on the horizontal branch
and on the blue side of the upper red giant branch.  Across
the diagram, the mean absolute residual is 1.3$\sigma$.

\begin{figure}
\plotone{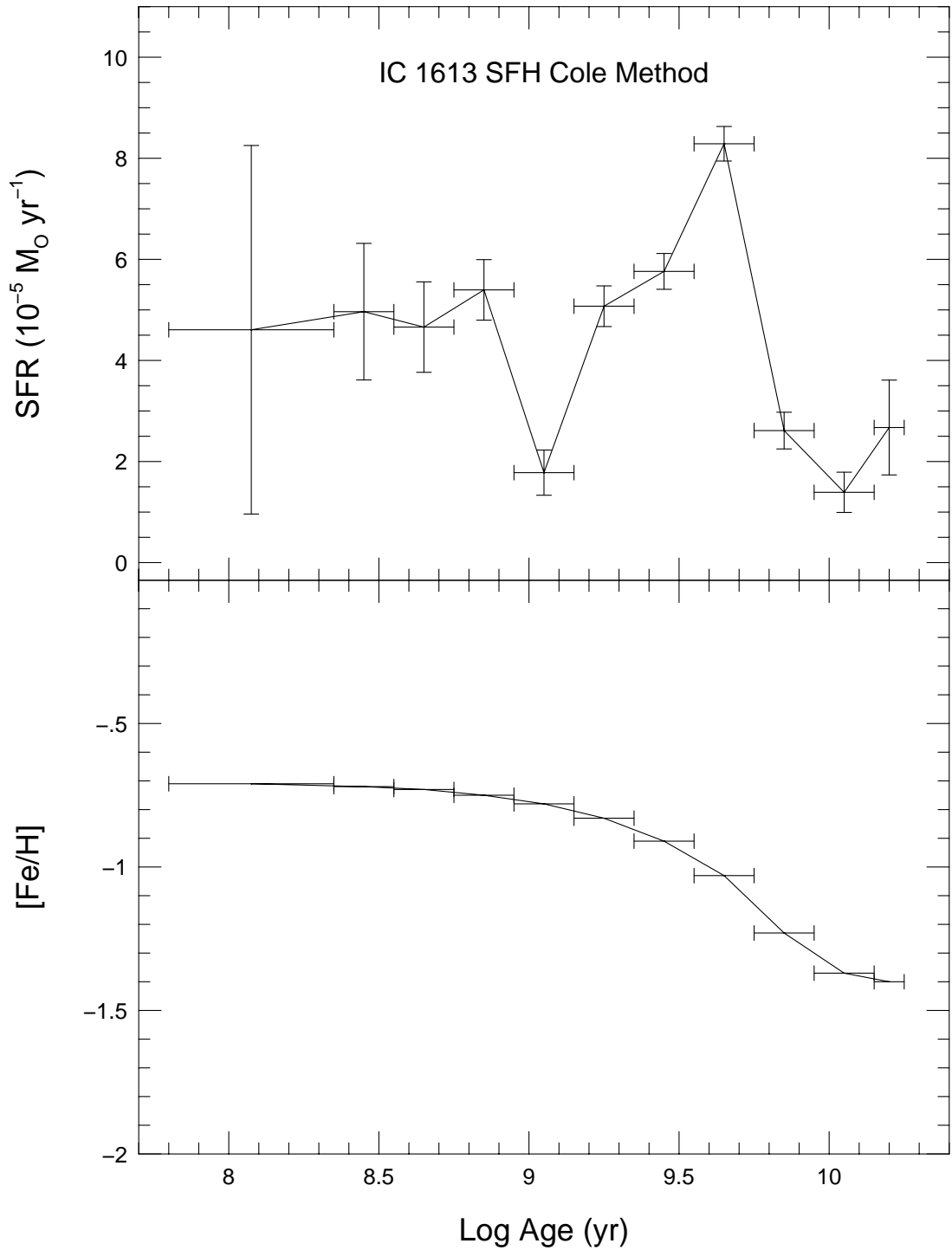}
\caption{ Plot of SFH and metallicity evolution
of IC~1613 derived via Cole method.  Note that the age-metallicity
relationship is an input for the models and not a direct result
of the solution.  Also note that the time axis is logarithmic
because logarithmic age binning was used.
}
\label{ac}
\end{figure}

Although the isochrones are spaced by 0.05 dex in age,
a more informative picture of the SFH is obtained if the
age binning is coarser.  Age bins 0.2 dex wide were adopted,
except for the youngest bin, which extends from
7.80 $\leq$ log (t/yr) $\leq$ 8.35.  The derived SFH, and the
assumed AMR, are shown in Figure~\ref{ac}.
The error bars on the SFH show the 1$\sigma$ errors on the
SFR in each age bin, considered in isolation.

The derived SFH is in broad agreement with that proposed
for the main body by C99, who suggested roughly constant
star-formation over the lifetime of the galaxy, with a
decline in the last 0.5 Gyr.  Our deeper data and more
detailed analysis suggest that the SFR
began quiescently, increased by a factor of three roughly 5 Gyr
ago, and has been declining slowly ever since.  Nearly half
of all stars were formed during the time period from 1.4--6 Gyr ago.

\subsection{The Dolphin approach}

The SFH was determined using the techniques
described by Dolphin (1997, 2002).  The underlying principle of this
approach is to determine the distance, extinction, SFH, and chemical
enrichment evolution most likely to produce the observed CMD.  
The solution was then derived according to the prescription given by
Dolphin (2002) over the CMD section 19.0 $\le$ V $\le$ 27.0, 
18.0 $\le$ I $\le$ 26.5, and $-$0.5 $\le$ V$-$I $\le$ 3.0.  
The bright limits were set by saturation; the faint
limits were set by where contamination from blended faint stars became
significant.  Throughout the solution, an IMF slope of $-$1.30, binary
fraction of 0.35, and flat secondary mass function were assumed.

Determination of the age and metallicity resolution of the solution
was made by balancing the fit improvement possible with
high-resolution solutions with the lower uncertainties given by
low-resolution solutions.  The balance was determined by minimizing
the ``maximum acceptable fit" (Dolphin 2002), as resolution was varied.
The optimal resolution was determined to be 0.1 dex in both age and
metallicity.

The best fit was measured at (m$-$M)$_0$ $=$ 24.25 and A$_V$ $=$ 0.05.  
Fitting the
fit parameter as a function of distance and extinction to a surface
and using the maximum acceptable fit to determine uncertainties, the
minima and uncertainties were measured to be (m$-$M)$_0$ $=$ 24.27 
$\pm$ 0.03 and A$_V$ $=$ 0.06 $\pm$ 0.03.  This value of the distance
modulus is in excellent agreement with that derived by D01 
((m$-$M)$_0$ $=$ 24.31 $\pm$ 0.06).

\begin{figure}	
\plotone{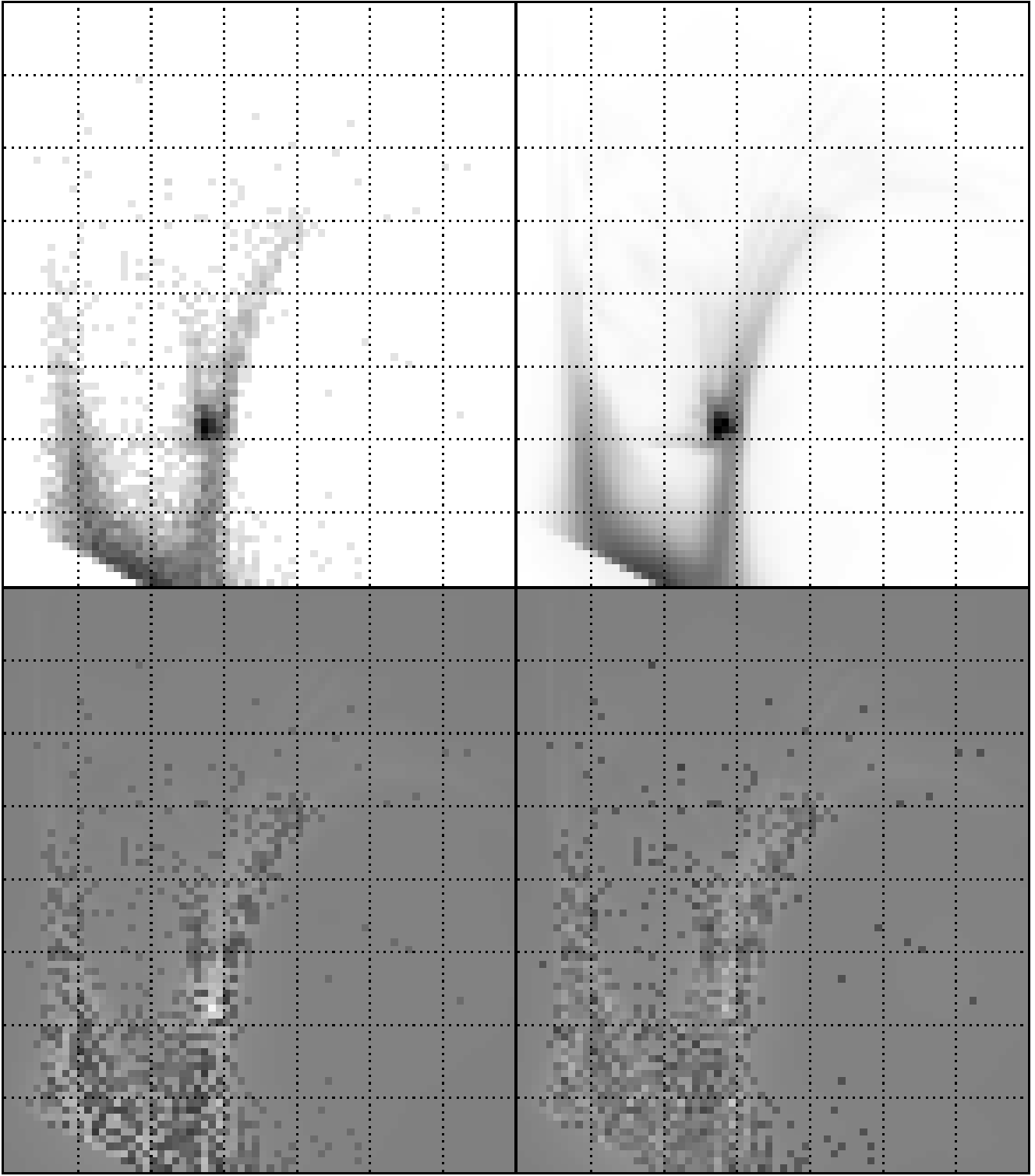}						     
\caption{High resolution Hess diagrams showing the comparison
between the observed and modeled CMDs: 
(upper left) observations;
(upper right) best-fit synthetic CMD; 
(lower left) residuals (light $=$ more observed; dark $=$ more synthetic);
(lower right) residuals in terms of sigma, the darkest shades
corresponding to a 3 $\sigma$ difference.
}
\label{adsyn}
\end{figure}

Figure~\ref{adsyn} shows a comparison of the observed and synthetic CMDs.  
The only significant differences are that the red clump is tighter and
the RGB broader in the synthetic CMD.  All points in the CMD were fit
with $\le$ 3$\sigma$ errors, and the overall quality of the best fit was 
4$\sigma$ from ideal.  This indicates a good, though not perfect, fit.

The largest problem with the fit appears to be the blue helium burning
sequence extending vertically up from the red clump.  Because the
models struggled to fit this feature adequately, it was partially
filled in using an extremely metal-poor red giant branch.  One could
certainly redo the solution with this part of the CMD omitted, but we
do not believe that the recovered SFRs were
significantly affected because the total number of RGB stars appears
to be correct.  (This is the case because the number of stars on the
RGB below the red clump is strongly constrained.)  A second problem,
also affecting the RGB, is that the shape of the theoretical RGBs does
not exactly match that of the observed data (see Figure 5), thus 
causing the model
RGB to be smeared slightly so that all observed points can be
accounted for.  Again, the number of RGB stars is constrained by the
lower RGB, and the effect on the recovered SFH is minimal.

\begin{figure}		
\plotone{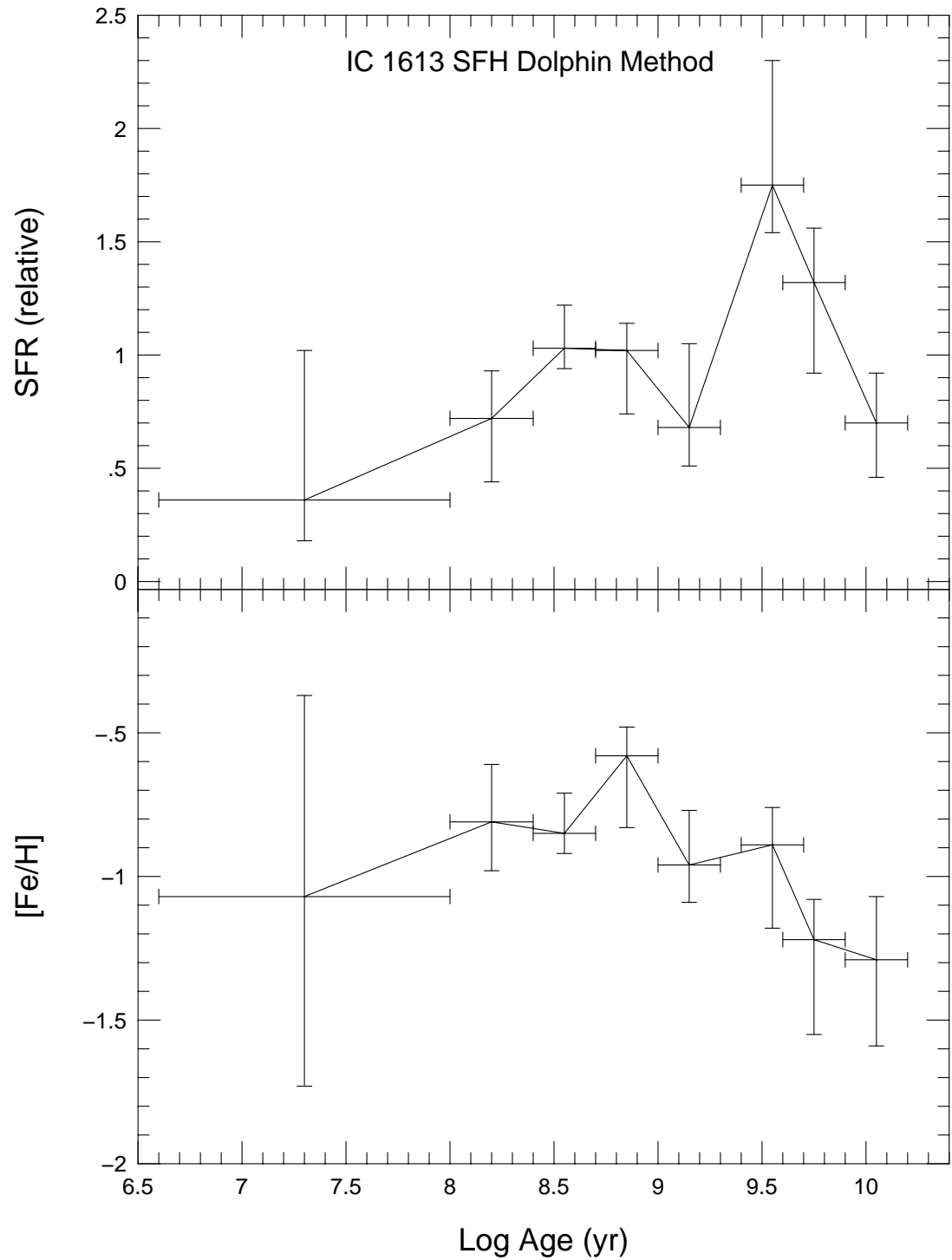}						     
\caption{Plot of SFH and metallicity evolution
of IC~1613 derived via Dolphin method.
Error bars derived as described in text.
}
\label{ad}
\end{figure}

The SFH and chemical enrichment history were also
measured, and are shown in Figure~\ref{ad}, binned by 0.3
dex in age.  The SFRs are relative values, normalized
to a mean value of 1.  The mean metallicity (weighted by mass of stars
formed, rather than number of stars present today) was measured to be
[Fe/H] $=$ $-$1.20 $\pm$ 0.20.  Error bars are not symmetric, as the best fit
(rather than the average of acceptable fits) is shown.  Uncertainties
were measured from two sources: scatter as distance and extinction
were moved within the acceptable fit range and scatter as measured
from Monte Carlo tests, in which synthetic populations were created
and measured.  Because of the highly-correlated errors in the SFRs
at various ages and metallicities, this composite
measurement of uncertainty provides more accurate estimates of the
uncertainties of these quantities than does searching around the
maximum acceptable fit.

The main feature seen in the SFH is an extended
event from $\sim$ 2 Gyr ago until 5 -- 10 Gyr ago.  While there has been star
formation since that event (a significant amount coming 0.5 Gyr ago),
the bulk of the stars in this region of IC 1613 come from the earlier
age.  Although Dolphin et al. (2001) found RR Lyraes in this field,
the ancient ($\ge$10 Gyr) SFR was well below the lifetime
average.

The metallicity enrichment appears to have been primarily during the
extended star formation episode.  Older than 10 Gyr, we find a mean
metallicity of [Fe/H] $=$ $-$1.39 $\pm$ 0.24 dex; by 1 Gyr ago this had
increased to [Fe/H] $=$ $-$0.63 $\pm$ 0.21.  
We see very little enrichment since
then, as the average metallicity of all stars younger than 1 Gyr is
[Fe/H] $=$ $-$0.77 $\pm$ 0.19.  It should be noted that no assumptions 
regarding the AMR were made in the solution; metallicities
are determined entirely from the populations required to fit the CMD.

\section{Comparison of Three Star Formation Histories}

\subsection{The Agreement Between the Different Methods}

\begin{figure}	
\plotone{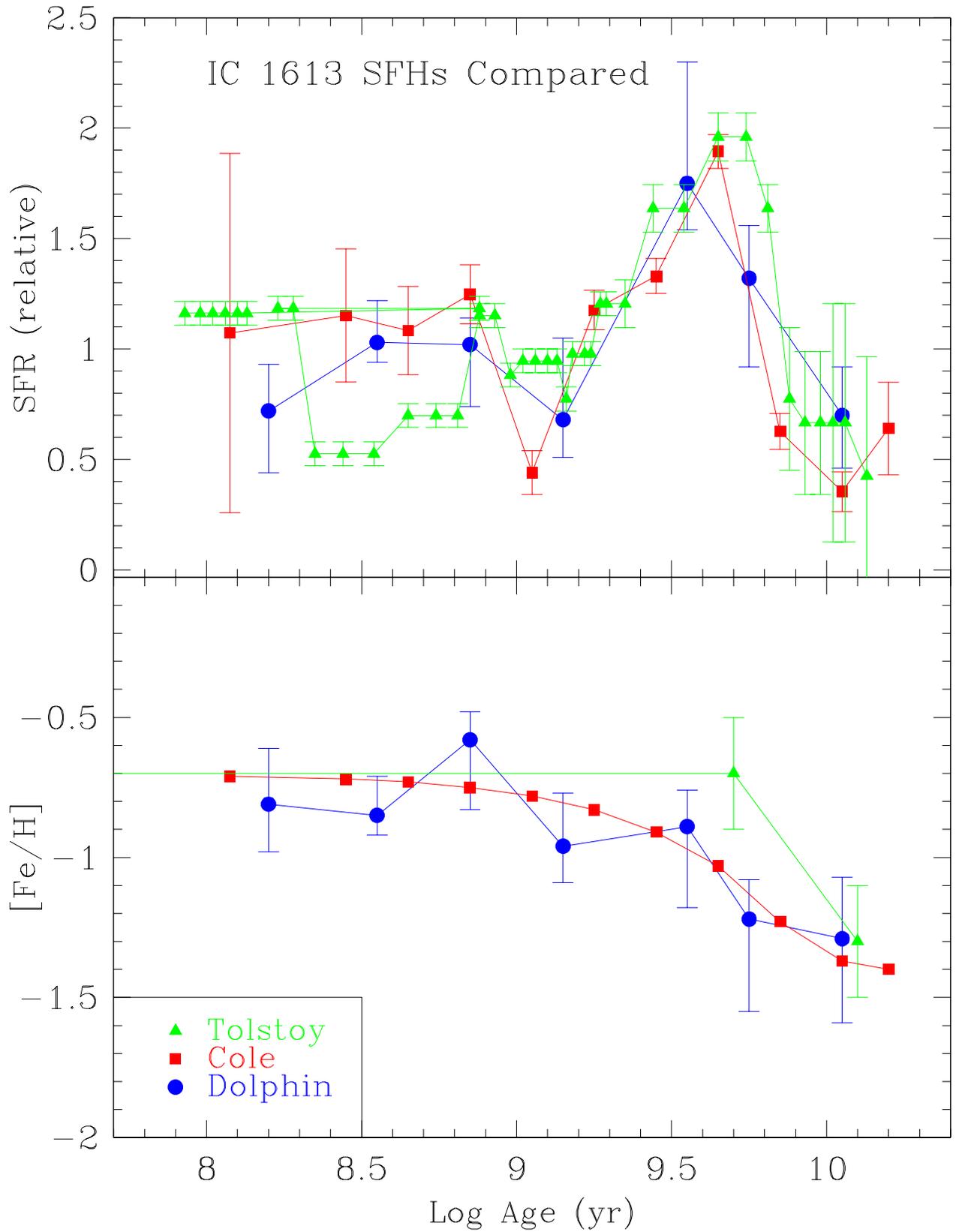}							     
\caption{Comparison of SFHs for
IC~1613 derived via three different methods.  Note that 
time axis is plotted logarithmically.  Reasonably good agreement
is found between all three methods.   
}
\label{comp}
\end{figure}

\begin{figure}	
\plotone{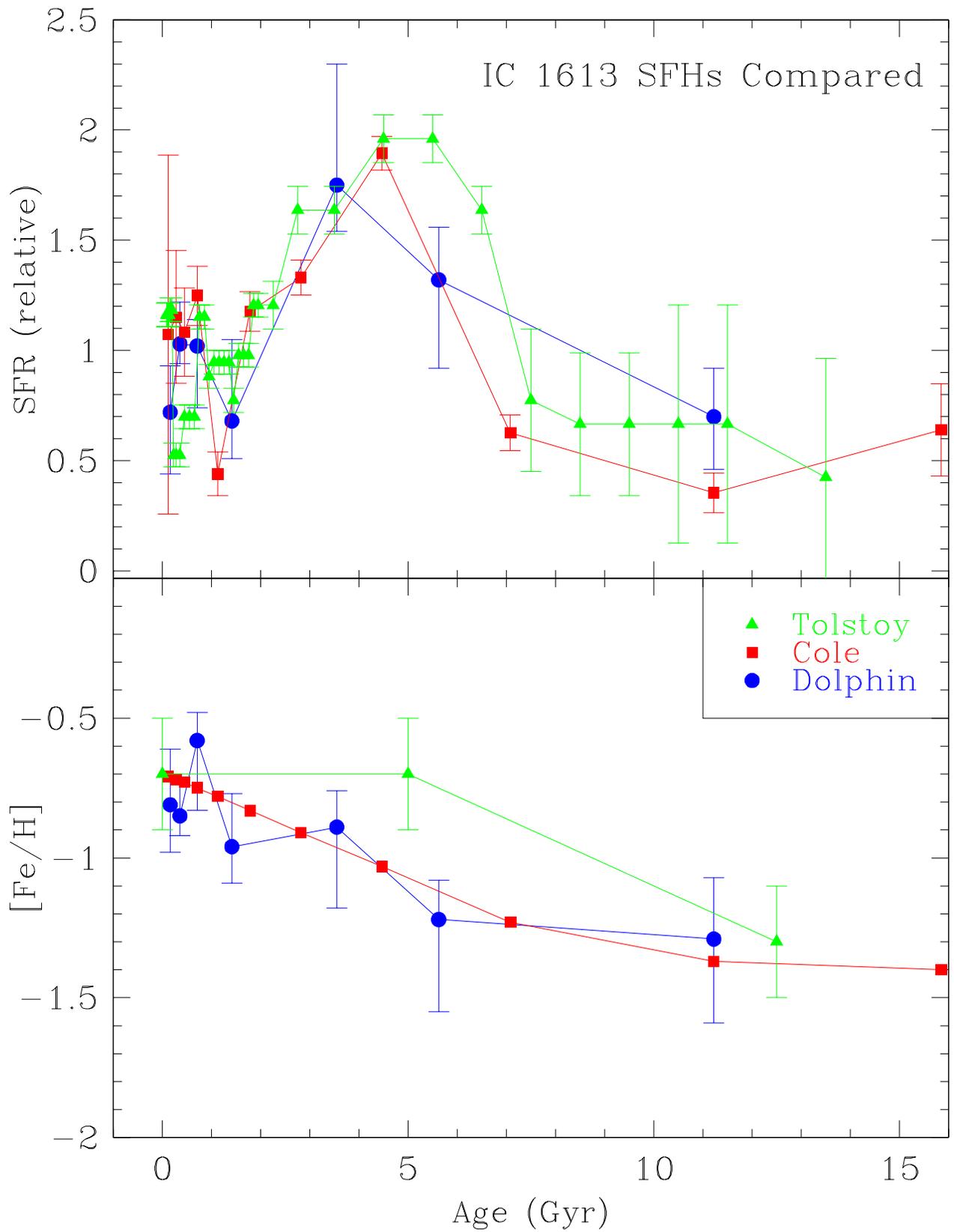}							     
\caption{Comparison of SFHs for
IC~1613 derived via three different methods as in Figure~\ref{comp},
but here the time axis is plotted linearly.  
Note the enhanced levels of SFR between 3 and 6 Gyr ago which appear in
all three models. 
}
\label{comp2}
\end{figure}

Figure~\ref{comp} shows a direct comparison of the best fit solutions
for the SFHs and chemical enrichment histories for IC~1613 derived
via the three different methods.  Figure~\ref{comp2} shows the same
comparison, only with the time axis plotted linearly instead of
logarithmically.  We find the agreement between the three methods
rather striking and remarkable considering the large differences
between the methods.  In particular, all three methods find an
enhanced SFR, at roughly the same magnitude
(factor of 3), over roughly the same period (from 3 to 6 Gyr ago).
The SFR at early and late times is relatively low, but 
significantly not zero.  Thus, IC~1613 appears to have been
creating stars over the entire history of the universe, without
very large ($\ge$ factor of 10) variations in SFR (subject to
our time resolution constraints).
Additionally, all three methods were driven to AMRs 
which are in excellent agreement.

Although the solutions do show general agreement, there are
clearly differences in the details.  Some of the differences may be 
artificial due to different choices of bin sizes in time.
The most notable difference appears in 
the last billion years, where the observational
time resolution is the greatest; the Tolstoy method finds
a relatively lower SFR over the period of 0.2 to 0.6 Gyr.
Overall, the differences between the models
appear to be consistent with the sizes of the estimated errors,
suggesting that the internal error estimates are, for the most
part, reasonable.

\subsection{The CMD Features Which Determine the Solution}
 
It is important to understand the CMD morphology that 
drives fitting algorithms, whether manual or automated,
to the solutions at which they arrive.  The primary 
empirical driver of the solutions found here is the 
large number of stars fainter than I $\approx$ 26.3,
with a wide color extent from 0.1 $\lesssim$ V$-$I 
$\lesssim$ 0.9.  
Such large numbers of stars are only produced if the SFR
was higher at some point in the past than it is now.  
This pedestal represents the main-sequence turnoffs
and subgiant branches of the dominant stellar populations
in IC~1613.  It was tentatively detected near the limit
of the data in C99, although crowding made it a low
reliability feature.  Because of the 
evolutionary timescales, main-sequence stars must be
the primary inhabitants of this region of the CMD, 
although some contribution of subgiants is present,
especially if there is a large old, metal-poor population.

It could be that the color extent
of the pedestal is largely responsible for the small differences
in ancient SFH derived by the three methods.  This is because
of the strong influence of metallicity on the color
of the main-sequence for low-mass stars, and to a lesser
extent because the presence of unresolved binary stars
tends to smear out stellar sequences to the red.  A significant
role is also played by the treatment of incompleteness,
which can shift stars significantly in the color-magnitude plane.
Although the differences are small, it is interesting to 
understand the origins of these differences. 

For the Tolstoy method (no interpolation in metallicity 
between stellar evolution tracks, no binary
stars, and an increasing with time but otherwise unrestricted
AMR), a relatively high metallicity 
($Z$ = 0.004) and old age (4--6 Gyr) are
necessary to match the red side of the pedestal. 
Where younger stars are present, they are assumed to be
of similar (or higher) metallicity -- they are brighter,
but not much bluer (see Figure~\ref{etmod}).  This leaves
the blue half of the pedestal to be filled in by the 
unevolved main-sequence of a much younger ($\sim$10$^8$ yr)
population.

In the case of isochrone interpolation, binary stars,
and an assumed AMR (Cole method), the age of the stars which
populate the pedestal is determined by the AMR.
Because of the contribution of binary stars,
the intrinsic sequence that matches the data can be bluer
than if no binaries were assumed.  The color is primarily
controlled by the metallicity, and so the age follows.
The smearing effect of binary stars and the interpolation
in metallicity allow for a more uniform population of the
blue half of the pedestal with younger and more metal-rich
stars.

In the case where isochrones are interpolated, binary stars
are included, and the AMR is explicitly
solved for by the automated fitting routine (Dolphin method), the code
is free to select ages and metallicities that make the
best fit to the pedestal.  Thus the reddest part of the
pedestal is fit by an old, metal-poor population, the middle
part is fit by a more metal-rich population at $\approx$3 Gyr,
and the bluest part is matched to a slightly more metal-poor, slightly 
younger population.  This is the inverse case to the previous one:
the color extent of the pedestal determines the AMR
(for stars older than $\approx$1 Gyr).

Choices propagate through the CMD to RGB color and width,
red clump and HB color, magnitude and morphology.  Unfortunately, 
the models are unreliable here (Bertelli \& Nasi 2001; D'Antona 2002;
Schiavon et al.\ 2002), so it is difficult to decide how
much astrophysical (as opposed to mathematical) weight to place
on the differences.  

\subsection{A Self-Consistent Solution?}

In principle, it is possible to derive an AMR from the SFH.
Is the AMR derived in this manner consistent with that derived 
from directly from the CMD?
Since, in the Dolphin solution, the metallicity is a free parameter, 
and thus, the chemical enrichment law is derived independently of
the SFH, it is important to determine if
the solution is self-consistent.  That is, has the minimization
found a chemical enrichment law which follows naturally from
the SFH, or does the solution show chemical enrichment
and star formation proceeding in an uncoupled manner?

In order to derive an AMR from the SFH, several assumptions are required.  
For example, by integrating the SFH, assuming constant IMF
and constant yield, using the instant (or nearly instant) recycling 
approximation, it is possible to derive a chemical enrichment law.  
Some of these assumptions are reasonably
constrained, while others are not.  The biggest uncertainty
is the history of the gas content of the galaxy.
It would be most convenient to model the evolution of the galaxy 
as a ``simple closed box'' (e.g., Searle \& Sargent 1972),
but this model is known to fail for the majority of dwarf irregular
galaxies in the sense that either mass inflow or outflow (or both)
is required (Matteucci \& Chiosi 1983).  However, since we have
no independent information concerning the history of the gas
content, we assume a simple closed box model evolution in
making a comparison calculation.  Using the present day gas mass 
of 8.2 $\times$ 10$^7$ M$_{\sun}$ (Lake \& Skillman 1989), a stellar
mass of 1.0 $\times$ 10$^8$ M$_{\sun}$ (based on an assumption of M/L(B) $=$ 1.4 
from the model calculated in \S 7.3), and a gas phase oxygen abundance 
of 6.8 $\times$ 10$^{-5}$, results in an oxygen yield of 8.4 $\times$ 10$^{-5}$, 
or 17\% of the solar oxygen abundance.  This ``effective yield''
is relatively low when compared to theoretical values of the 
oxygen yield (e.g., Maeder 1992), but this is typical of the Local 
Group dwarf irregulars (see discussion in Kennicutt \& Skillman 2001). 
With this value of the yield, it is then possible to calculate
a chemical enrichment history for a closed box model with the
calculated star formation rate as a function of time.

\begin{figure}	
\plotone{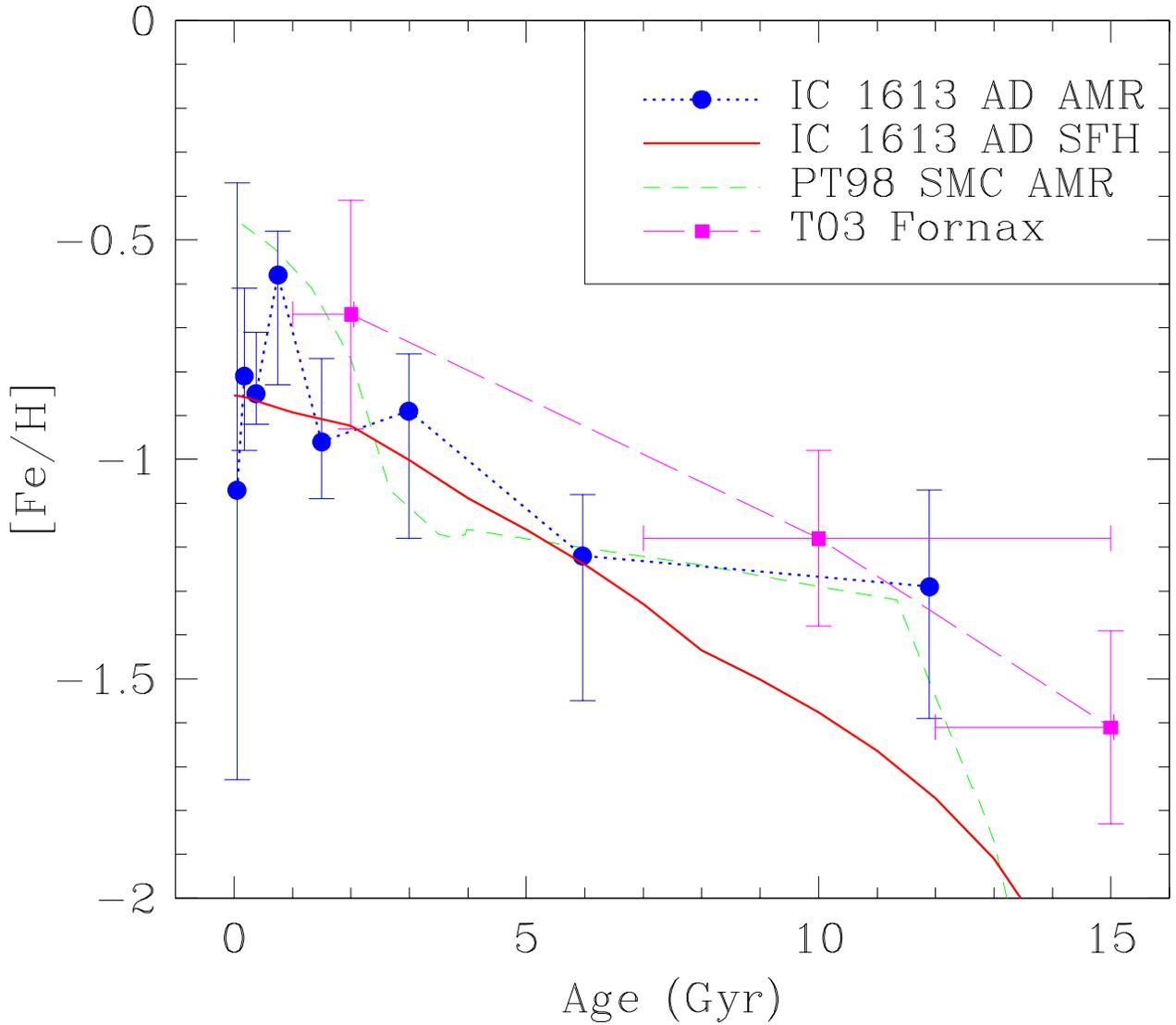}							     
\caption{A comparison of the metal enrichment histories for IC~1613 derived
from the Dolphin method CMD analysis (blue filled circles with errorbars connected by
dotted line) with that derived by assuming that the enrichment is directly 
tied to the SFR (solid red line) assuming a closed box model evolution.  
Note the good agreement between the two independent determinations over most
of the range in age.  The divergence at early ages could be due to prompt
initial enrichment or the fact that our time resolution at earliest times 
is very poor.  Also plotted are the age metallicity relationship
for the SMC (short dashed green; Pagel \& Tautvai\v{s}ien\.e 1998) and 
spectroscopic abundances for three stars in the Fornax dSph (magenta filled 
squares with error bars connected by long dashed line; Tolstoy et al.\ 2003).  
Note that all three galaxies, although of different morphological type, 
have similar age-metallicity relationships.
}
\label{modzfh}
\end{figure}

In Figure~\ref{modzfh} we have plotted the metal enrichment history
of IC~1613 derived from the CMD via the Dolphin method 
and compared it to a metal enrichment history calculated from the 
SFH derived via the Dolphin method assuming a simple 
closed box model with constant yield.
As can be seen in Figure~\ref{modzfh},
the agreement between the two is quite good, with the possible exception
of the earliest ages.  This agreement provides another check on the
validity of the CMD analysis.

The small difference at the earliest ages is
in the sense that the model does not rise quite as quickly as the
CMD analysis predicts.   There are several possible explanations for
this difference (if real).
Certainly our CMD analysis does not provide a strong constraint
on the SFR at the earliest epoch, so it could be that there is early
star formation that is not found in our solution.  For example, we have
assumed constant star formation during the interval up to 8 Gyr ago;
if we assume a declining SFR during that interval, then the metallicity
rises more quickly. 
Alternatively, it could be that yields are not constant with time or metallicity
(e.g., there could have been an epoch of prompt early enrichment by a
population of stars with an unusual IMF; Larson 1998).
For comparison, we have included in Figure~\ref{modzfh} the age-metallicity
relationship for the SMC as derived by Pagel \& Tautvai\v{s}ien\.e (1998),
and that measured for the Fornax dSph from high resolution spectroscopy
of individual stars by Tolstoy et al.\ (2003). 
Note that the enrichment law for the SMC appears to fit better in the
sense of a steeper rise at early times.  This is a result of assuming
inflow (there is less gas to enrich at early times, so the gas is enriched
faster).  Clearly there are several uncertainties in 
understanding the nature of the enrichment in these dwarf galaxies. 
A significantly deeper CMD and spectroscopic metallicities will be
required to better constrain the various possibilities.

\subsection{Implications for CMD Fitting Methodology}

Despite significant differences in technique between the different
modeling approaches, the basic shape of the derived SFH
is robust.  At the broadest level, the SFHs
are insensitive to the details of the comparison.  This
phenomenon was also seen in the comparison of SFH models for the
Large Magellanic Cloud (Skillman \& Gallart 2003),
but is perhaps more surprising for the case of IC~1613, where
there is no clear indication for a distinct main-sequence turnoff,
as there is in the LMC.

The impression one gets from this comparison is that all three
methods are well constrained by the observations and driven to
similar results in calculating the SFHs.  Can this result be
generalized?  There are even more methods of deriving SFHs from
CMDs, so this exercise, by no means, covers the spectrum.
For example, several methods rely on a much coarser binning
within the CMD.  Additionally, this exercise has only been
conducted on one observed CMD.  There could be CMDs representing
different SFHs where the results of the different methods do
depart significantly.
Finally, all three methods used the same stellar evolution
models as input, so uncertainties in the input models are not
being tested.  If the models change significantly in the future,
then. of course, one would expect different SFHs as a result.

In sum, we can conservatively say that this exercise supports the 
claim (e.g., Dolphin 2002) that high quality CMDs (with the MS sampled 
down to M$_V$ $=$ $+$2 or fainter) will provide sufficient information
to place strong constraints on the SFH of a galaxy.  
The main goal now is to obtain similarly high quality CMDs
for other dwarf irregular galaxies and other galaxies
sampling different morphological types and environments. 

\section{The Evolution of Star-Formation and Metal Enrichment in IC~1613}

\subsection{The Outer Field}

Have the deep observations presented here significantly
improved our knowledge of the SFH and chemical enrichment of IC~1613?  
Hodge (1989) provided a sketch of the ``probable'' SFH and AMR of IC~1613,
consistent with its known properties, showing a constant SFR
and a gradual enrichment of heavy elements.
Mateo (1998), based on the observations of Freedman (1988) and
Saha et al.\ (1992), also shows a relatively constant SFR, with
a slight decrease at intermediate ages.
The SFH presented by Grebel (1999), based on Freedman (1988) and
Cook \& Aaronson (1988) is nearly identical
to that of Mateo, but shows constant metallicity
over the entire age of the galaxy.
It would appear that all of the previous SFHs for IC~1613 agree in showing a
lack of strong bursts of star formation (within the
limits of the time resolution possible).

From a comparative study of the masses, blue luminosities, and H$\alpha$
fluxes of a sample of irregular galaxies, Gallagher, Hunter, \& Tutukov (1984)
concluded that dwarf irregular galaxies had nearly constant SFRs
over their lifetimes.  Early studies of the upper portions of the CMDs of
nearby irregular galaxies supported this conclusion (e.g., Tosi et al.\ 1991).  
The present observations are also consistent with this conclusion.   
However, the new observations reveal, for the first time, the enhanced SFR 
over a 3 - 4 Gyr period at intermediate ages and give us a first real look at
the chemical enrichment in a dIrr as a function of time.  

We will consider the increase in SFR at intermediate ages relative to 
the earliest ages to be real.  It is true that, with the present uncertainties, 
the SFH history of IC~1613 is marginally
consistent with constant star formation over the history
of the universe.  However, we feel that three independent analyses indicating
a heightened SFR at intermediate ages is interesting (possibly surprising) and
should not be ignored.  In \S 6 we will compare this result with what is
seen in other, well studied dwarf galaxies.  In \S 7 we will
discuss this within the context of modern theoretical expectations
for the evolution of dwarf galaxies. 

The chemical enrichment history of IC~1613 is now known in better 
detail than ever before.  The derived chemical enrichment is in
good agreement with that expected for a system showing nearly 
constant star formation, and is similar to that of other well studied
dwarf galaxies.
We have included in Figure~\ref{modzfh} the age-metallicity
relationship for the SMC as derived by Pagel \& Tautvai\v{s}ien\.e (1998),
and that measured for the Fornax dSph from high resolution stellar spectroscopy 
by Tolstoy et al.\ (2003).  Despite the
differences in morphologies between the three galaxies (isolated dwarf irregular,
interacting irregular galaxy, and dSph), all three galaxies have very similar
AMRs.  Perhaps this is not too surprising
as all three galaxies are apparently dominated by star formation at 
intermediate ages.

\subsection{Comparison Between the Inner and Outer Fields of IC~1613}

The outer field is 6$\farcm$7 southwest of the central C99 field, and 8$\farcm$2
southwest of the geometric center of IC~1613.  At 1.8 kpc projected
distance from the center, the field is $\approx$2.4 scale lengths
out in the disk (Hodge et al.\ 1991), well within the edge of the H{\small I}
distribution (Lake \& Skillman 1989).  Should we expect a difference in CMD
morphology, due to radial gradients in age or metallicity, between
the two fields?  Baade noted early on that the blue, irregularly
distributed stars in IC~1613 are embedded in a much larger, regular
sheet of red stars; clearly some kind of gradient in stellar population
parameters is present.  Our outer field is just beyond the ``blue''
limit of IC~1613 as quoted by Baade (1963),
and more than halfway to the edge of the red sheet.  This 
color difference should be reflected in the CMD morphologies.

To the limiting magnitude achieved here, we are most sensitive to
age gradients over the past $\sim 10^9$ yr, which will be
manifest in the ratio of main-sequence and supergiant
stars relative to red giants and red clump stars.  Metallicity
differences should be reflected in color shifts, following the
well known rule of thumb that line blanketing in stellar
atmospheres pushes stars to the red at high metallicity.

\begin{figure}
\plotone{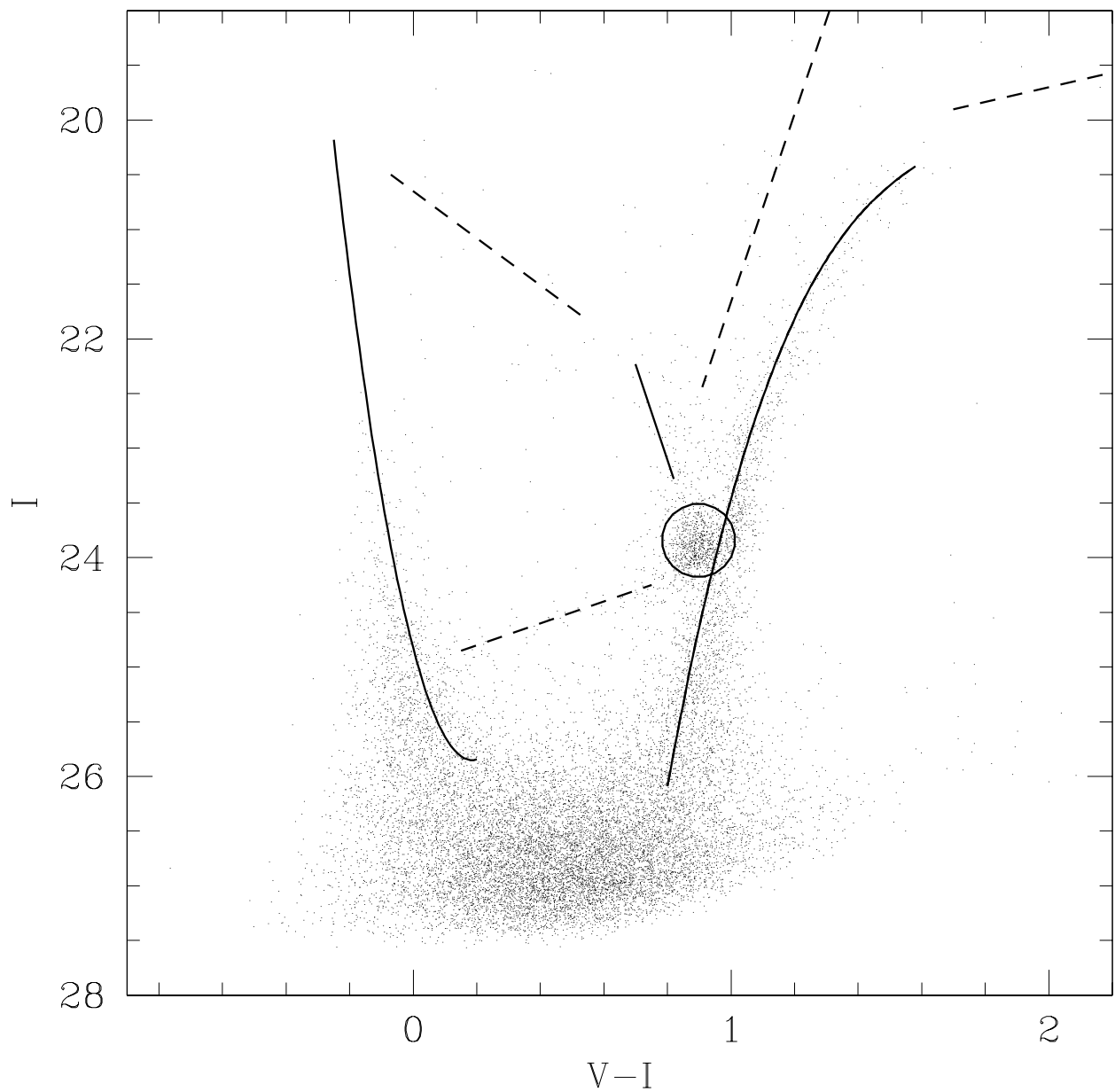}
\caption{Fiducial sequences of the central field overplotted on
the CMD of the outer field.}
\label{inout}
\end{figure}

A direct comparison of the two fields (e.g., by a differential
Hess diagram) is difficult because of the quite different limiting
magnitudes and crowding levels in the images.  In Figure~\ref{inout}
we show the CMD of the outer field (compare to Figure~\ref{footprint}) with 
fiducial sequences derived from the C99 field overplotted.
Solid lines mark the high density features in the CMD: the 
main-sequence, RGB, red clump, and ``vertical red clump'' 
(intermediate-mass blue loop stars).  Dashed lines delineate the
less populous features: horizontal branch, blue and red supergiants,
and thermally-pulsing AGB.

There are no significant color differences between the
principal sequences of Figure~\ref{inout}.  Below I $\approx$25,
both the RGB and the MS fiducials of the C99 field appear
to deviate from the outer field sequences.  However, the
differences are small and in opposite directions; they are
interpreted as due to the larger errors and increased crowding
in the C99 data.  The agreement between upper RGBs is striking:
if any metallicity gradient is present, a finely-tuned age
gradient must be invoked to cancel out the effect.  It is
unlikely that the dominant population in the C99 field is
very much younger than the outer field, given 1) the relatively
young age of most stars in the outer field (Figure~\ref{comp2}), and
2) the approximately steady-state history derived by C99 
for the central field.  The similar colors of the intermediate-age
and old stellar populations in both fields suggests that they
are not very different from each other, at least over most
of their history.  Further support for this view derives
from the near constancy with radius of the carbon:M-star
ratio (Albert, Demers, \& Kunkel 2000), indicating a comparable 
proportion of
intermediate-age stars at both small and large radii.

The relative strengths of the various features, however,
vary significantly between the two fields.  Table 1 
lists the relative numbers of stars in various evolutionary
phases; the uncertainties in the ratios are derived from
Poisson statistics on the number counts. 
The difference in surface brightnesses in photographic
$J$ band is taken from Hodge et al.\ (1991) as 2.4$\pm$0.5 mag arcsec$^{-2}$,
or a factor of 9$\pm$3 in flux.  Because photographic $J$ is 
blue-sensitive, it is not surprising that the unresolved $J$-light
is most closely tracked by the bluest resolved stars.  There
is a clear difference in number density ratios 
between the young stars and the old stars. 

The available information suggests that as the global
activity level in IC~1613 declined after the broad peak
$\sim$ 2--7~Gyr ago, the star-formation rate in the
central regions stayed high longer, and has maintained
an average level some 30--40\% higher than the outer field.
Star formation remains much higher still in the off-center
complex of H{\small II} regions that coincides with the
peak of the H{\small I} column density (Lake \& Skillman 1989).
In the outer field, star formation has slowed but not
stopped, as attested by the presence of a few small
H{\small II} regions even at this radius (Hodge, Lee, \& Gurwell
1990).

\section{A Comparison of IC~1613 to Other Dwarf Galaxies}
 
\subsection{Comparison to Other Dwarf Irregular Galaxies} 

\subsubsection{Star Formation Histories}
 
Although detailed SFH studies exist for other dwarf irregular
galaxies, no dwarf irregular galaxies at distances beyond the 
Magellanic Clouds have been observed to the depth of the present 
study of IC~1613 (M$_V$ $\simeq$ $+$3.4).
Dolphin (2002) has shown that the CMD needs to be sampled down
to at least M$_V$ $=$ $+$2 for meaningful constraints on the 
ancient and intermediate age SFH,
and most studies of dwarf irregular galaxies fall short of this.
For example, the pioneering study of NGC~6822 by Gallart et al.\ 
(1996) used ground based photometry with a limiting M$_V$ 
of $-$1.  This has only recently been surpassed by the HST 
observations of Wyder (2001), with a limiting M$_V$ $=$ $+$1
for those observations.  The HST observations of WLM reported by
Dolphin (2001c) reached to M$_V$ $\simeq$ $+$2 and the HST observations
of the ``transition'' galaxy LGS~3 reached to M$_V$ $\simeq$ $+$2.5
(Miller et al.\ 2001).  HST observations of the  relatively nearby
Phoenix dwarf, which today is classified as a transition galaxy but 
presumably would have been classified as a dwarf irregular galaxy 
until only very recently ($\sim$ 100 Myr ago), has been observed down 
to M$_V$ $\simeq$ $+$4.0 (Holtzman, Smith, and Grillmair 2000).
By comparison, our HST study of Leo~A (Tolstoy et al.\
1998) reached to M$_V$ $=$ $+$1, and this has only recently been 
improved to M$_V$ $\approx$ $+$1.5 by Schulte-Ladbeck et al.\ (2002).
Thus, truly secure SFRs for the ancient populations of isolated dwarf
irregular galaxies are quite scarce.  

Note that one of the fundamental conclusions of this study,
that star formation was relatively low at ages older
than 7 Gyr compared to intermediate ages, is only possible with the
present depth of photometry.  In order to be able to determine
whether this is a general feature of low mass dwarf irregular
galaxies or particular to IC~1613, CMDs of similar depth are
needed for a large sample of dwarf irregular galaxies.  At present,
this will only be possible with the HST for Local Group galaxies.

IC 1613 is not the first dwarf irregular galaxy for which it
has been suggested that star formation has occurred predominantly 
at intermediate ages. 
The SMC is also known to have a SFH which
is dominated by star formation at intermediate ages (Dolphin et al.\ 2001c
and references therein).  From the study of a single field in
the SMC, Dolphin et al.\ (2001c) find a slow rise in the SFR
from the earliest ages to a peak at 7 Myr and then a steep
decline to the present.  This is very similar to that for IC~1613,
with the exception of a higher SFR in the last Gyr for IC~1613.
It is interesting that these two galaxies, with presumably
very different evolutionary histories (IC~1613 evolving in
isolation and the SMC being strongly influenced by both the
LMC and the Galaxy), have very similar star formation
and chemical enrichment histories.  Note, however, that the
changes in the SFR in the Magellanic Clouds are often associated with
the tidal effects of a close encounter or close passage to the Milky Way
galaxy (although the SMC and LMC have different SFHs,
see discussions in Gardiner \& Hatzidimitriou 1992 and
Olszewski et al.\ 1996).

There are also galaxies for which it has been suggested that most 
of the stars have been formed relatively recently.
Our earlier study of Leo~A (Tolstoy et al.\ 1998) suggested
that the majority of the star formation had occurred in the
last 2 Gyr.  Dolphin et al.\ (2002) discovered  
RR Lyraes in Leo~A, and thus the presence of very early star 
formation, but converting the number of RR Lyrae stars to an
early SFR is very uncertain. 
Interestingly, the
Schulte-Ladbeck et al.\ (2002) study of Leo~A concludes
that the CMD is consistent with either similar SFRs at 
earliest and intermediate ages or decreasing SFR from 
earliest to intermediate ages, but not increasing SFR
from earliest to intermediate ages as we see here for
IC~1613.  It would be interesting to model the newer HST
Leo~A observations with the techniques described in this paper 
to see if the same conclusion is derived. 
A similar SFH is found for Sextans~A.  From a relatively deep
(M$_V$ $\simeq$ $+$2.0) CMD, Dolphin et al.\ (2003) find that 
while there is evidence for very old stars in Sextans~A, the
SFR at intermediate ages (3 - 10 Gyr) was quite low, and the
SFR has been the highest in the last 2 Gyr. 

However, of those dwarf irregular and transition galaxies with 
relatively deep CMDs, the SFH history found in IC~1613 does not 
appear to be universal.  In WLM, Dolphin (2000c)
finds a relatively high SFR at the earliest times, which 
is never reached again in the history of the galaxy.
In LGS~3, Miller et al.\ (2001) find a SFR which is highest
in the first few Gyr, and then lower by at least a factor of
10 thereafter.  The Phoenix dwarf galaxy shows a relatively high
SFR in the earliest few Gyr followed by fluctuations but showing
an overall decrease in SFH.  These large differences in star 
formation histories probably point to different environmental
influences, but identifying which environmental effects are
dominant remains a challenge.

\subsubsection{Radial Gradients in Star Formation Histories}

Studying population gradients in dwarf irregular galaxies does
not require the same depth of photometry as constraining the
oldest SFRs.  In fact, it was in IC~1613 that Baade (1963) identified 
the underlying sheet of Population II red giant stars that 
covers twice the area on the sky that the blue stars do.
It appears that all of the dwarf irregular galaxies
that have been studied in sufficient detail do show radial
gradients in their stellar populations (e.g., SMC, Gardiner \& 
Hatzidimitriou 1992; WLM, Minniti \& Zijlstra 1997; 
GR~8, Dohm-Palmer et al.\ 1998; LGS~3, Miller et al.\ 2001;
Leo~A, Dolphin et al.\ 2002).  This effect is seen to be especially
strong in the blue compact galaxies (e.g., UGC 6456, Lynds et al.\
1998; Schulte-Ladbeck, Crone, \& Hopp, 1998; Aloisi et al.\ 2001;
Tosi et al.\ 2001; Dolphin et al.\ 2001a; Cannon et al.\ 2003).  
It is interesting
that Harbeck et al.\ (2001) find the same thing for dSphs;
however, they interpret the majority of the dSph cases as  
metallicity gradients and an age gradient only for Carina. 

Do the strong population gradients observed in low mass star forming
galaxies imply that current star formation is confined exclusively
to the inner parts of these galaxies or can star formation proceed
in the outer parts, but at a greatly reduced level?  In the case
of IC~1613 (and some other low mass dwarf irregulars) a large fraction
of the HI gas is located outside of the optically defined galaxy.
Gallagher \& Hunter (1984) introduced the notion of ``usable gas'';
under the assumption that massive star formation is a threshold 
phenomenon, gas that lies below the critical threshold would not
be usable gas, i.e., not available for star formation.  Most of 
our knowledge of star formation in the outer parts of galaxies
has come from studies of their HII region distributions 
(but see recent study of the distribution of high surface brightness 
areas by Parodi \& Binggeli 2003).  However, detailed studies of the 
stellar populations allow us a much clearer picture of the star 
formation patterns as functions of  both time and radius.
Unfortunately, the present study represents just a small fraction of
the area of IC~1613.  It would be interesting to study the stellar
population gradient over the entire face of IC~1613.

The strong age gradients mean that scale lengths must be changing
with time.  Of course, this needs to be properly accounted for in 
comparison of nearby galaxies with their potential progenitors at
intermediate redshift (e.g., Barton \& van Zee 2001; Pisano et al.\ 2001).

\subsection{Comparison to Dwarf Spheroidal Galaxies}

The currently favored $\Lambda$CDM models for galaxy formation afford
a significant role for dwarf galaxies.  Dwarf galaxies are assumed to
trace early star formation in the Universe and galaxy evolution
until the present.
Due, in large part, to attempts to understand the possible evolutionary
connections between the dwarfs with negligible or extremely low present
SFRs (the dSph galaxies) and the dwarfs with obvious signs of present 
star formation (e.g., dIrrs, blue compact dwarfs, HII galaxies), many 
theorists are turning their
attention to the problem of dwarf galaxy evolution.  Environmental effects
are turning out to be a key parameter (e.g., van den Bergh 1994a,b;
Klypin et al.\ 1999; Moore et al.\ 1999; Gnedin 2000b; Mayer et al.\ 2001a,b;
Carraro et al.\ 2001).

We have produced the deepest CMD of an isolated dwarf irregular
galaxy.  Given the density-morphology relationship in
the Local Group, the dIrr galaxies are at much greater distances
than the dSphs, and, thus, have correspondingly shallower CMDs.
As a result, it has been difficult to make direct comparisons
between the CMDs of dIrr galaxies and dSph galaxies.  
Even though our CMD for IC~1613 is still not as deep as the typical 
study of a dSph MW companion, we can now attempt such a comparison.

\begin{figure}	
\plotone{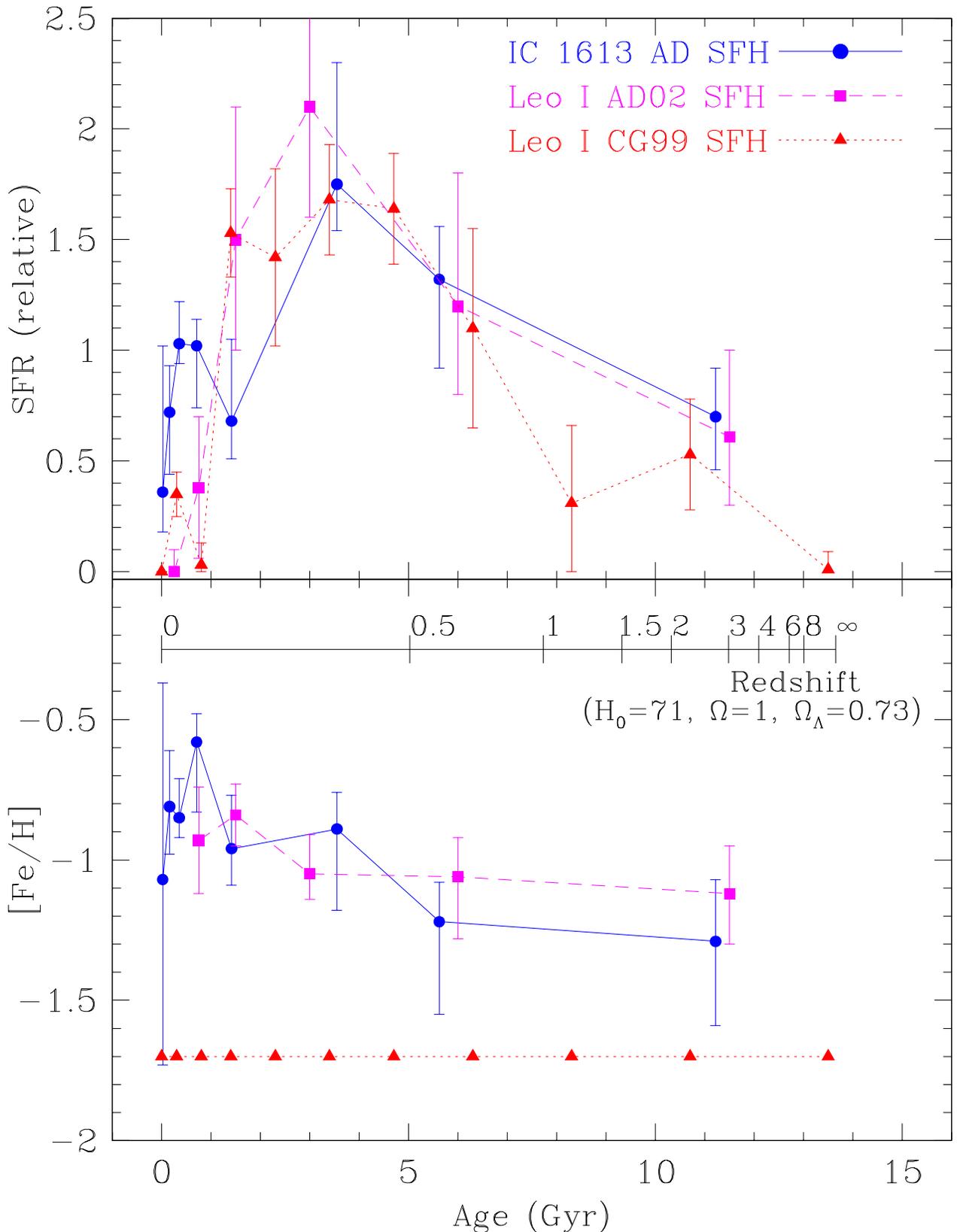}							     
\caption{Comparison of SFHs and metal enrichment histories
for IC~1613 (derived via the Dolphin method) and the dSph Leo I (derived via the 
Dolphin method, Dolphin 2002; and by Gallart et al.\ 1999a).
Note that the two galaxies have nearly identical star formation and metal 
enrichment histories when analyzed in an identical fashion.
A timeline comparing redshift to real time has been added for the noted
cosmology for convenience.  Note that the bulk of the star formation and 
chemical enrichment has occurred at $z$ $<$ 1.0.
}
\label{compleoi}
\end{figure}

To first order, star formation in IC~1613 appears to have occurred
predominantly at intermediate ages.  This is also 
characteristic of several of the outer MW dSph satellites
(specifically Carina, Fornax, Leo~II and Leo~I, see introduction).
A particularly interesting comparison is between IC~1613 and
Leo I; both appear to be dominated by star formation at the same 
intermediate ages (Gallart et al.\ 1999a; Dolphin 2002).
In Figure~\ref{compleoi} we compare the Dolphin SFH and AMR solutions 
for IC~1613 with the  Dolphin (2002) solutions for Leo~I.  
We also add the Gallart et al.\ (1999a) solutions for Leo~I.
Figure~\ref{compleoi} shows that the SFHs and AMRs for IC~1613 and 
Leo~I, when derived via identical methodology, are nearly identical. 
Figure~\ref{compleoi} also shows very good agreement between the 
SFHs derived by two different methods, even though the derived
AMRs are quite different (the main difference most likely arising because 
the upper red giant branch, which is quite sensitive to metallicity,
is not used in the methodology of Gallart et al.\ 1999a).  

One possible interpretation of Figure~\ref{compleoi} is that,
absent the youngest stars, is it possible that there are no
differences between the stellar populations of isolated dIrr and 
dSphs which are more distant from their parent galaxies.  The implication
is that dIrr and some dSph galaxies have similar progenitors,
and that the differences which we see today are due to 
environmental influences during the lifetimes of the galaxies
which allow one type of galaxy
to retain its gas and form stars up the present and another not.
Certainly the morphological census of the Local Group has
evolved with time. 

Do the similar SFHs and AMRs for IC~1613 and Leo~I support the proposal 
by Lin \& Faber (1983) that dSphs can be formed by ram pressure
stripping of dIrrs?  At the time, Lin \& Faber estimated that an
ambient number density of 10$^{-6}$ cm$^{-3}$ would be required for
stripping the gas from a dIrr over a Hubble time and inferred the
presence of such gas through indirect arguments.  Recently,  FUSE and
CHANDRA observations have provided evidence
for extended coronal gas surrounding either the Milky 
Way galaxy or the Local Group with a density of $\sim$ 10$^{-6}$ cm$^{-3}$
(Nicastro et al.\ 2002; 2003).  Additionally, Stanimirovic et al.\ (2002) 
estimate surprisingly high densities of $\sim$ 10$^{-4}$ cm$^{-3}$ for 
the Galactic halo via observations of pressure confined clouds in the 
Magellanic Stream.  Thus, there does appear to be material with the
potential for stripping the gas from Local Group dwarf galaxies.
Indeed, the HI cloud associated with the Phoenix dwarf (Young \& Lo 1997;
St.\ Germain et al.\ 1999; Irwin \& Tolstoy 2002) may be direct evidence
of a stripping event having occurred in the last 100 Myr
(Holtzman et al.\ 2000).

The main problem with drawing a direct evolutionary link between IC~1613
and Leo~I would appear to be the total number of stars.  With an M$_V$ 
$=$ $-$14.7, IC~1613 is much more luminous than Leo~I (M$_V$ $=$ $-$11.9).
While small differences in luminosity could be explained by fading of
the dSph after truncation of star formation, the nearly 3 magnitude
difference would be difficult to account for.
Although the V-band central surface brightnesses are comparable (22.8 $\pm$ 0.3
for IC~1613 versus 22.4 $\pm$ 0.3 for Leo~I, Mateo 1998), the optical scale 
lengths are significantly different ($\sim$ 700 pc for IC~1613 and 
$\sim$ 130 pc for Leo~I).  Given that Leo~I is a dark matter dominated
system, it does not seem likely that ram pressure stripping of the HI
gas would result in a significantly greater optical scale length.
(Indeed, Faber \& Lin calculated that IC~1613 was too large to evolve into
a present day dSph.)  A possible solution to this could be that only the
central region of Leo~I has a smaller scale length and that there is an
underlying larger scale length stellar distribution (perhaps as suggested by
Stoehr et al.\ 2002).  In sum, the difference in luminosity between IC~1613
and Leo~I is another manifestation of the bimodal nature of the 
metallicity-luminosity relationships for dwarf galaxies (Mateo 1998;
Grebel et al.\ 2003), which remains as a key hurdle for the proposal
of converting dSphs to dIrrs via ram pressure stripping.

\section{Impact on Our Understanding of Galaxy Evolution}

Low-mass dwarf irregular galaxies provide an important testing ground
for several fundamental questions about galaxy evolution  
and cosmology.  Given our SFH for IC~1613,
we now consider the implications under the assumption that
IC~1613 is typical of dwarf irregular galaxies which are
evolving primarily as isolated galaxies in low relatively
low density environments.

\subsection{Evidence for Suppression of Early Star Formation by
Photoionization?}

An especially interesting question in this regard is the effect of
reionization on the suppression of dwarf galaxy formation.
Originally discussed within the context of explaining the 
absorbers responsible for the Ly-$\alpha$ forest (Ikeuchi
1986; Rees 1986), later papers realized the potentially
important effects of early heating of the ISM on the evolution 
of dwarf galaxies (Efstathiou 1992; Babul \& Rees 1992; 
Chiba \& Nath 1994;
Quinn, Katz, \& Efstathiou 1996; Thoul \& Weinberg 1996;
Kepner, Babul, \& Spergel 1997;
Barkana \& Loeb 1999; Bullock et al.\ 2000).
With a rotation curve maximum of 25 km s$^{-1}$, IC~1613 is
clearly in the regime of galaxies which should be susceptible
to this effect.  The model of Babul \& Rees (1992) specifically 
predicts a gap or a fallow period
of star formation from the time of reionization ($z$ $\ge$ 6, Fan et al.\ 2002;
or $z$ $=$ 17 $\pm$5, Spergel et al.\ 2003) to the 
time when the UV background dropped sufficiently for the gas in 
low mass haloes to cool (which they estimate to be about $z$ $\sim$ 1).  
A redshift timeline has been added to Figure~\ref{compleoi} so that one can 
look at the SFH as a function of both real time and redshift.  Although the
time resolution is course, it appears that the heightened SFR in
IC 1613 started between $z$ of 0.5 and 1.
It could be argued that the SFH of IC 1613 is exactly what one
expects from an isolated galaxy with roughly 10$^9$ M$_{\sun}$.  
It might have formed as a virialized structure sometime before reionization,
and started to make stars then.
After reionization, the high UV background prevented further infall and 
kept the gas within IC~1613 from cooling rapidly, so that it never formed too
many stars until star formation in M31 and the Milky Way settled down enough 
to drop the UV background, at a redshift roughly between 1 and 2.  It then took 
some time for that gas to cool, until finally at z $\approx$ 0.5
the star-formation ramped up to a peak and slowly declined later.
This is not ``delayed galaxy formation,'' as we do see evidence for
very old stars in IC~1613 (e.g., RR Lyrae stars), but formation of a 
dwarf in which much of the star formation was deferred.  

Within the collapse and dissipation model of galaxy formation (White \& Rees
1978), the effects of reionization on the evolution of dwarf galaxies is 
twofold (see discussion in Benson et al.\ 2002a and references therein).
The ionizing background heats the IGM which suppresses the collapse of 
low mass structures.  The ionizing background can also heat the ISM in 
low mass haloes which have already formed and reduce the rate of 
radiative cooling by reducing the number of neutral atoms.  The first effect
has a strong influence on the low end of the galaxy luminosity function.
The photoioinzing background will reduce the fraction of gas which collapses
with the dark matter especially in systems which are less massive 
than the ``filtering mass'' (Gnedin 2000b).  Benson et al.\ (2002a)
point out that the formation history is also an important parameter in
determining the final gas content of a dark matter halo.  In order to
solve the perceived problem of the mismatch between the theoretically 
expected and observed low end of the galaxy luminosity function, much
attention has been given to the effects of the suppression of the 
formation of dwarf galaxies (e.g., Bullock et al.\ 2000; Chiu, Gnedin, 
\& Ostriker 2001; Somerville 2002; Benson et al.\ 2002b, 2003; Tassis et al.\
2003).  However, these studies generally concentrate on the luminosity 
function and star formation histories of individual galaxies are not
explored in depth.  

The second effect is relevant to the question of suppressed star formation 
at intermediate ages. 
Can star formation in dwarf galaxies which have already formed before
reionization be suppressed by the photoionizing
background and then recommence at redshifts of 0.5 and lower?  
Babul \& Rees (1992) point to the 
fast decline in the UV background between $z$ = 2 and the present and 
estimate that new star formation would precipitate in dwarfs at 
$z$ $\le$ 1.  Calculating a specific SFH depends strongly on the
influence of both external (the photoionizing background) and internal
(stellar feedback) variables (Tassis et al.\ 2003).  The evidence of an 
environmental dependence of the low end of the galaxy luminosity function 
(Tully et al.\ 2002; Trentham \& Hodgkin 2002) may imply a very large 
range of SFHs due simply to environmental differences
on the effects of photoionization (Benson et al.\ 2003).
 
Could the low SFR in IC~1613 at early ages indicated
by our CMD analysis be the signature of suppression of 
star formation by the photoionizing background?  The effect could 
be environmentally dependent (i.e., for dwarf galaxies close to 
large galaxies, the effects of the UV radiation from the neighbor
galaxies could be more important than the realtively uniform UV 
background from distant sources), so a statistically
significant sample of SFHs for dwarf irregular galaxies will
be needed for any definitive statements.  However, it is
interesting that the bulk of the star formation in IC~1613 
occurs at intermediate ages, which may well be after much of
the gas in the central parts of the galaxy has had time to 
recombine and cool.  It is also interesting that the star formation
and chemical enrichment histories for IC~1613 are similar to those of the 
SMC and Leo~I.  These three galaxies show very different morphologies,
so why should their evolutionary characteristics be so similar?
This would be possible if their evolution is driven by external
forces and if these galaxies shared relatively similar environments.
The extreme cases of Leo~A (Tolstoy et al.\ 1998) and Sextans~A 
(Dolphin et al.\ 2003), in which it appears that star formation has been 
suppressed until $z$ $\approx$ 0.2 -- 0.3, may present even stronger 
challenges.  Note that while Sextans~A is nearly identical to IC~1613
in terms of luminosity and rotation curve maximum (Skillman et al.\ 1988) 
Leo~A is an order of magnitude fainter and shows almost no rotation at 
all (Young \& Lo 1996). 

However, it should be noted that the evolutionary characteristics of IC~1613 
are not universal.  Early enhancements in SFR are detected in several of 
the dwarf galaxies in the Local Group.  Our suggestion that we are able
to detect the effects of reionization in the SFHs of
the nearby dwarfs may raise as many questions as it answers.  
The UV background is just one of many factors which can influence the
SFH of a dwarf galaxy, and it may be the dominant factor for only a
fraction of dwarf galaxies.
At this point it would be very useful for the galaxy evolution modelers
to present compilations of the SFHs for the dwarf galaxies in their
simulations.  As the quality of the CMDs in the observational studies
increases, and thus the time resolution improves, detailed comparisons
between observational SFHs and theoretical simulations may prove
to better constrain ideas about the relative importance of different
effects on dwarf galaxy evolution.  

\subsection{Enrichment of the IGM by Outflows from Dwarf Galaxies?}

It is currently thought that the bulk of the
intergalactic medium (IGM) was chemically enriched above roughly
0.1\% of the solar value at least by $z$ $=$ 3 and possibly as
early as $z = 5$ (Songaila 2001; Pettini et al.\ 2003; 
Boksenberg, Sargent, \& Rauch 2003; and references therein).
Because of the lower masses of dwarf galaxies, they are a
natural source to consider for the formation and ejection of
the metals that are found in the IGM (e.g, Dekel \& Silk 1986; 
Nath \& Trentham 1997; Efstathiou 2000; Garnett 2002).  
The SFH for IC~1613 indicates that it may have produced the bulk of 
its metals at intermediate
ages (i.e., 3 - 6 Gyr ago).  This is not the correct time frame
for the observed enrichment of the IGM.  Assuming the currently
favored $\Lambda$CDM model with $\Omega$ $=$ 1.0, 
$\Omega_{\Lambda}$ $=$ 0.73, and H$_0$ $=$ 71 km s$^{-1}$
(Freedman et al.\ 2001; Spergel et al.\ 2003), a
redshift of 3 corresponds to a lookback time of 
11.5 Gyr (for an age of the universe of 13.7 Gyr).
Thus, the slow time scale for the production of elements for present 
day dwarf irregular galaxies does not fit in
with what is needed for the rapid enrichment of the IGM.

Perhaps this should be clarified.  This does not rule out prompt
enrichment by the first structures to form (e.g., Tegmark et al.\ 1997) 
or ``pre-galactic outflows'' (e.g., Madau, Ferrara, \& Rees 2001;
Mori, Ferraro, \& Madau 2002).
In fact, there are many classes of models which successfully
explain the early enrichment (cf.\ Aguirre et al.\ 2001b).
Notably, the recognition that outflows appear to be universal in 
Lyman break galaxies at high redshift (Pettini et al.\ 2001) allows the
early enrichment to be conducted entirely by larger ($\sim$ 10$^{10}$ M$_{\sun}$)
galaxies (Aguirre et al.\ 2001a).  Currently, a great deal of attention 
has focused on the supermassive stars thought to form only
at the earliest epochs (e.g., Oh et al.\ 2001).
The only scenario which we cast doubt on
is that the present day dwarf galaxies are responsible for
the bulk of the prompt enrichment of the IGM via star formation with
a ``normal'' IMF.

\subsection{Dwarf Galaxies and the Faint Blue Galaxy Population}

Wyder (2001) explored whether analogs to the Local Group 
dwarf irregular NGC~6822 could be contributing to the faint
blue galaxy excess.  Based on WFPC2 images that reached the
level of the red clump, he concluded that NGC~6822 was likely never
bright enough for its $z$ = 0.5 analogs to contribute to 
current redshift surveys.  However, his derivation of the
history of NGC~6822 lacked age resolution, due to the 
relatively shallow (M$_V$ $<$ $+$1)  limiting magnitude of 
the data.  Thus an extreme burst of star-formation in 
NGC~6822 could temporarily brighten it enough to qualify
as a faint blue galaxy, while remaining consistent with
the observed CMD.

Using our deeper data for IC~1613, and motivated by the
evidence for a significantly higher star-formation rate
during the epoch corresponding to redshifts 0.3--0.6, we
can estimate the likelihood that IC~1613 analogs are 
represented in large-scale galaxy surveys.  Following
Wyder (2001), we adopt an updated version of the PEGASE.2
evolutionary synthesis code (Fioc \& Rocca-Volmerange 1997)
to model the evolution of IC~1613's integrated magnitudes
and colors.  We take a rough average of the three SFHs
presented in Section 3 as our baseline input model for the
synthesis code.  Because the extinction to IC~1613 is small,
it can be neglected to first order.  The PEGASE.2 output
is then renormalized to match the observed absolute B magnitude
of IC~1613, M$_B$ = $-$14.6.  A very basic check on the modeling
procedure is provided by the output integrated color, B$-$V = 0.53.
This is in good agreement with the observed B$-$V = 0.6 $\pm$ 0.1.

Our baseline model for IC~1613 has been gradually brightening
for most of its history, with a steeper increase after its
era of enhanced star formation roughly 3--6~Gyr ago, and a 
small subsequent decline.  It reached
its brightest absolute B magnitude M$_B$ = $-$15 during this
episode.  For reasonable $\Lambda$CDM cosmologies, this corresponds
to a redshift $z$ $\approx$ 0.4.  Using the formulae for 
cosmological distance modulus (e.g., Peebles 1993), and appropriate
spectral windowing to account for the K-correction, the baseline
model for IC~1613 would have apparent magnitudes B $\approx$ 27.5 and 30
at redshifts $z$ = 0.5 and 1, respectively.  This is far too faint
to be included in the present generation of redshift surveys.

If IC~1613 had a history of starbursts, it could have been much 
brighter and bluer in the past than the relatively smoothly evolving baseline
would predict.  We modify our baseline model by replacing
the star-formation rate between ages of 2.5 and 7 Gyr, i.e., we excise
the era of enhanced star-formation rate.  We replace the removed 
portion with an equal mass of stars, but we force them to form during
an 0.1~Gyr interval centered at 5.0~Gyr.  This lookback time would
correspond to a redshift of $z$ = 0.5 for a distant analog of IC~1613.
The timescale is chosen to be a realistic estimate for the burst 
lifetimes in nearby dwarf galaxies (cf.\ Cannon et al.\ 2003).  
The resulting burst has an
amplitude of 20 over the long-term average star-formation rate.
This history produces a synthesis model that is 0.7 mag brighter 
(in rest-frame B) than the baseline model at $z$ = 0.5, when normalized
to reproduce the present-day M$_B$ and B$-$V.  The relatively small
effect on the integrated magnitudes of the model galaxy can be traced
to the long era of low star-formation which precedes the burst, diluting
the impact of the numerous massive, young stars which are injected
into the model 5~Gyr ago.  The smaller K-correction implied by the
larger fraction of UV-bright stars in the bursting model makes it 
relatively even brighter than the baseline model, but still at 
the very faint limit of current redshift surveys (e.g., Lilly et al.\ 1995).

The perturbation in both color and magnitude dies down after less than 
10$^9$ yr, so that bursts have to be frequent and extreme in order
to add many IC~1613-like galaxies into the observed samples of faint 
blue galaxies.  Note that such models with   
high burst amplitudes and short durations then also require 
large numbers of quiescent objects (Tyson \& Scalo 1988).
However, there is no evidence in our data for
such an event (in agreement with the lack of older star clusters,
c.f., van den Bergh 2000b).  The expected errors on an individual data point
at the level of the main-sequence turnoff of a 5~Gyr old population
are comparable to the separation between isochrones aged 3 and 6~Gyr
at the distance of IC~1613 (see Figure 4).  However, the aggregate
distribution of data points are inconsistent with a dominant coeval
population smeared by photometric errors.  An instantaneous burst with 
an amplitude of 20 would produce a clear envelope to the subgiant branch
and a steep break in the main-sequence luminosity
function that are not observed.  This is reflected in the analyses
which led to our baseline model.   Somewhat smaller bursts, staggered
in time, could be hidden in the data, but these would have much
less effect on the integrated model colors.  We find it highly 
unlikely that analogs of IC~1613 have contributed to the excess
of faint blue galaxies in existing redshift surveys.
This is in agreement with the results of 
Carollo \& Lilly (2001), who found that a subsample of star-forming
galaxies in the redshift range 0.5 $<$ $z$ $<$ 1 have metallicities
more like present-day giant galaxies than today's dwarfs.

We can also ask whether the Sloan Digital Sky Survey (SDSS) will
be sensitive to analogs of IC~1613 at cosmologically interesting
distances.  York et al.\ (2000) give limiting magnitudes of 
$g^{\prime} \approx$ 23.3 for the SDSS, and our synthesis models show
roughly neutral (V~$-$~$g^{\prime}$) colors for IC~1613.  The absolute
V magnitude of IC~1613 is $-$15.2, and so it would be cataloged
by SDSS out to distance moduli $\mu _0$ $\lesssim$ 38.5, or distances
of $\approx$ 500~Mpc.  This corresponds to redshift $z$ $\approx$ 0.1.

\section{Summary and Conclusions}

We have taken deep images of an outlying field in the Local Group
dwarf irregular galaxy IC~1613 with the WFPC2 aboard the Hubble Space
Telescope in the standard broad-band F555W (V, 8~orbits) and
F814W (I, 16~orbits) filters.
We analyze the resulting color-magnitude diagram (CMD) and compare
it with CMDs created from theoretical stellar models using three
different methods to derive a SFH as well as
constrain the chemical evolution for IC~1613.

All three methods find an
enhanced SFR, at roughly the same magnitude
(factor of 3), over roughly the same period (from 3 to 6 Gyr ago).
Additionally, all three methods were driven to similar
age-metallicity relationships which
show an increase from [Fe/H] $\approx$ $-$1.3 at earliest
times to [Fe/H] $\approx$ $-$0.7 at present.
We investigate the age-metallicity relationship which is
derived independently of the SFH in order to
see if the two are self-consistent.
Good agreement is found between
the AMR which is derived from the CMD
analysis and that which can be inferred from the derived
SFH at all but the earliest ages (where our model SFH is
not well constrained).
The agreement between the three models and the self-consistency of
the derived chemical enrichment history support
the reality of the derived SFH of IC~1613 and, more generally,
are supportive of the practice of
constructing galaxy SFHs from CMDs.

A comparison of the newly observed outer field with an earlier
studied central field of IC~1613 shows that the SFR
in the outer field has been significantly depressed during the last Gyr.
This implies that the optical scale length of the galaxy has
been decreasing with time and that comparison of galaxies at
intermediate redshift with present day galaxies should take
this effect into account.

Comparing the CMD of the outer field of IC~1613 with CMDs of
Milky Way dSph companions, we find strong similarities between
IC~1613 and the more distant dSph companions (Carina, Fornax,
Leo I, and Leo II) in that all are dominated by star formation
at intermediate ages.  In particular, the SFH and AMR for IC~1613
and Leo I are indistinguishable.  This implies that dIrr galaxies
cannot be distinguished from dSphs by their intermediate
age stellar populations.  This may support scenarios of dwarf galaxy
evolution where dSph galaxies are produced via the stripping of
gas from dIrr galaxies.

This type of a SFH may also be evidence
for slower or suppressed early star formation in dwarf galaxies
due to photoionization after the reionization of the universe
by background radiation.  Theoretical models have suggested that
star formation will be suppressed until a redshift of $\sim$ 1,
and we find that most of the star formation in IC~1613 has
taken place after a redshift of 0.5.

Assuming that IC~1613 is typical of a dIrr evolving in isolation,
since most of the star formation occurs at intermediate
ages, ``normal'' star formation (i.e., with a universal IMF) in these 
dwarf systems cannot be responsible for
the fast chemical enrichment of the IGM which is seen at high redshift.
There is no evidence for any large amplitude bursts of star formation
in IC~1613, and we find it highly unlikely that analogs of IC~1613
have contributed to the excess of faint blue galaxies in existing
galaxy redshift surveys.

\acknowledgements

Support for this work was provided by NASA through grant GO-07496 
from the Space Telescope Science Institute, which is operated by
AURA, Inc., under NASA contract NAS5-26555.
EDS is grateful for partial support from a NASA LTSARP grant No. NAG5-9221
and the University of Minnesota.
EDS also gratefully acknowledges valuable discussions with Carme Gallart, 
Don Garnett, Martin Haehnelt, Rob Kennicutt, Henry Lee, Jerry Ostriker, 
Bernard Pagel, and Max Pettini, and the hospitality of the Institute of
Astronomy of the University of Cambridge during his sabbatical visit.
Martin Haenelt, Bernard Pagel, and an anonymous referee provided 
valuable comments on an earlier version of this manuscript.
ET gratefully acknowledges support from a fellowship of the Royal
Netherlands Academy of Arts and Sciences 
and JSG acknowledges partial funding from
the University of Wisconsin-Madison Graduate School.
This research has made use of NASA's Astrophysics Data System
Bibliographic Services and the NASA/IPAC Extragalactic Database
(NED), which is operated by the Jet Propulsion Laboratory, California
Institute of Technology, under contract with the National Aeronautics
and Space Administration.


\clearpage
\begin{table}
\caption{Stellar Density Ratios for Central vs.\ Outer Field}
\begin{tabular}{lccc}
\hline
Feature                         & N(central) & N(outer) & Ratio        \\
\hline
Main Sequence (V $\leq$ 24.5)   & 2023     & 240      & 8.4 $\pm$0.6 \\
Red Supergiants                 & 521      & 69       & 7.6 $\pm$1.0 \\
Blue Loop                       & 681      & 102      & 6.7 $\pm$0.7 \\
Red Clump (23.1 $<$ I $<$ 24.5) & 8333     & 1434     & 5.8 $\pm$0.2 \\
RGB (I $\leq$ 23.1)             & 1890     & 344      & 5.5 $\pm$0.3 \\

\hline
\end{tabular}
\end{table}
\clearpage

\clearpage

\clearpage

\clearpage

\clearpage

\newpage

\newpage

\end{document}